\documentclass[10pt,journal,compsoc]{IEEEtran}

\usepackage{courier}
\usepackage{amsmath}
\usepackage{amssymb}
\usepackage{amsfonts}
\usepackage{xcolor}
\usepackage{multirow}
\usepackage{float}
\usepackage{graphicx}
\usepackage{graphics}
\usepackage{subfigure}
\usepackage{float}
\usepackage{bigstrut}
\usepackage{booktabs}
\usepackage{bibentry}
\usepackage{balance}

\ifCLASSOPTIONcompsoc
  \usepackage[nocompress]{cite}
\else
  \usepackage{cite}
\fi
\ifCLASSINFOpdf
\else
\fi

\begin{document}

\title{Multi-behavioral Sequential Prediction with Recurrent Log-bilinear Model}

\author{Qiang~Liu,
Shu~Wu,~\IEEEmembership{Member,~IEEE,}
and Liang~Wang,~\IEEEmembership{Senior Member,~IEEE}
\IEEEcompsocitemizethanks{\IEEEcompsocthanksitem Qiang Liu, Shu Wu and Liang Wang are with the Center for Research on Intelligent Perception and Computing (CRIPAC), National Laboratory of Pattern Recognition (NLPR), Institute of Automation, Chinese Academy of Sciences (CASIA) and the University of Chinese Academy of Sciences (UCAS), Beijing, 100000, China.
\protect\\
E-mail: \{qiang.liu, shu.wu, wangliang\}@nlpr.ia.ac.cn.}

\thanks{}}

\markboth{Journal of \LaTeX\ Class Files,~Vol.~14, No.~8, August~2015}%
{Shell \MakeLowercase{\textit{et al.}}: Bare Demo of IEEEtran.cls for Computer Society Journals}

\IEEEtitleabstractindextext{%
\begin{abstract}
With the rapid growth of Internet applications, sequential prediction in collaborative filtering has become an emerging and crucial task. Given the behavioral history of a specific user, predicting his or her next choice plays a key role in improving various online services. Meanwhile, there are more and more scenarios with multiple types of behaviors, while existing works mainly study sequences with a single type of behavior. As a widely used approach, Markov chain based models are based on a strong independence assumption. As two classical neural network methods for modeling sequences, recurrent neural networks cannot well model short-term contexts, and the log-bilinear model is not suitable for long-term contexts. In this paper, we propose a Recurrent Log-BiLinear (RLBL) model. It can model multiple types of behaviors in historical sequences with behavior-specific transition matrices. RLBL applies a recurrent structure for modeling long-term contexts. It models several items in each hidden layer and employs position-specific transition matrices for modeling short-term contexts. Moreover, considering continuous time difference in behavioral history is a key factor for dynamic prediction, we further extend RLBL and replace position-specific transition matrices with time-specific transition matrices, and accordingly propose a Time-Aware Recurrent Log-BiLinear (TA-RLBL) model. Experimental results show that the proposed RLBL model and TA-RLBL model yield significant improvements over the competitive compared methods on three datasets, i.e., Movielens-1M dataset, Global Terrorism Database and Tmall dataset with different numbers of behavior types.
\end{abstract}

\begin{IEEEkeywords}
Collaborative filtering, sequential prediction, multi-behavior, recurrent log-bilinear.
\end{IEEEkeywords}}

\maketitle

\IEEEdisplaynontitleabstractindextext

%
\IEEEpeerreviewmaketitle

\section{Introduction}
\IEEEPARstart{N}{owadays}, Collaborative Filtering (CF) \cite{koren2011advances} plays an important role in a large number of applications, e.g., recommender systems, information retrieval and social network analysis. Conventional CF methods focus on modeling users¡¯ preference based on their historical choices of items and always ignore the sequential information. It is reasonable to assume that user preferences change with his or her behavioral sequence. Meanwhile, rather than with merely one type of behaviors, e.g., purchasing in e-commerce and clicking on websites, there are many sequential scenarios with multiple types of behaviors towards items, e.g., clicking, purchasing, adding to favorites in e-commerce and downloading, using, uninstalling in app usage. Accordingly, it is necessary to model multi-behavioral sequences and collaboratively predict what a user will prefer next under a specific behavior. For instance, multiple types of behaviors, i.e., posting, sharing and commenting, on social media has been separately modeled and studied recently, which makes great contribution to user interest detection \cite{zhao2015improving}. Besides e-commerce and other Internet applications, multi-behavioral sequential prediction can be implemented for social good, such as predicting security events in a specific area \cite{liu2016strnn}\cite{wu2016sape} or predicting air quality \cite{yu2015air}.

\begin{figure}[tb]
\centering
\includegraphics[width=0.48\textwidth]{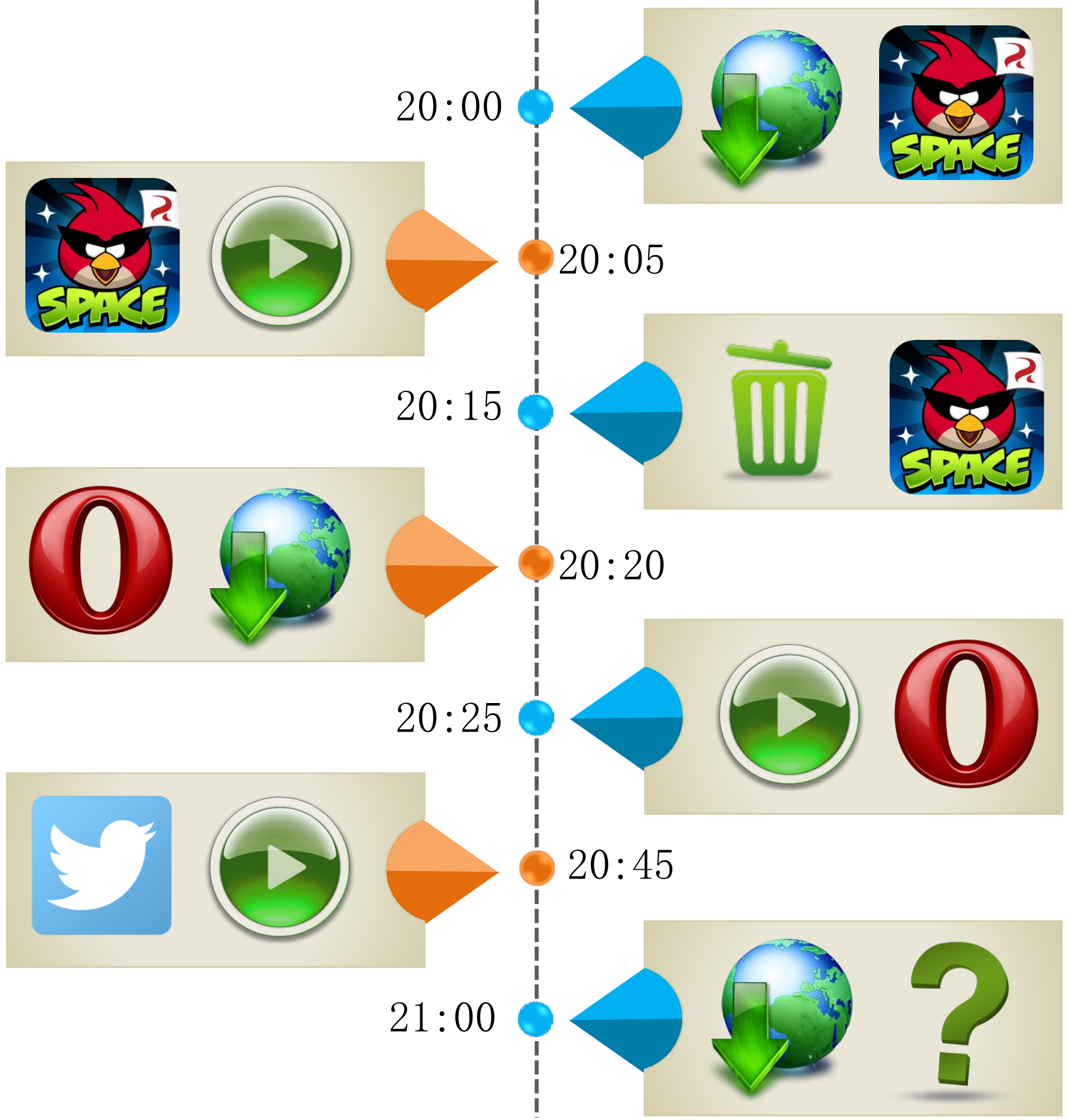}
\caption{Taking app usage prediction as an example of multi-behavioral sequential prediction. This example shows a user's behaviors towards apps in an hour, including downloading, using and uninstalling. We can predict what app the user is going to download or use next.}
\label{fig:Example}
\end{figure}

Nowadays, some efforts have been put into developing CF methods with sequential information \cite{campos2014time}\cite{liu2016strnn}\cite{rendle2010factorizing}\cite{wang2015learning}\cite{zhang2014sequential}. To the best of our knowledge, none of existing methods are designed for modeling sequences with multiple types of behaviors. And if we directly treat different behaviors towards one item as different elements in sequences, or simply ignore the differences among behaviors, conventional methods will have difficulty in revealing the correlations among behaviors and items. As shown in the example of app usage in Figure \ref{fig:Example}, different behaviors reveal users' different attitudes towards apps. Downloading and using means you may like the app, while uninstalling means you do not like the app and similar ones should not be recommended. So, it is essential to find a proper way to reveal the correlations among behaviors and items.

Moreover, existing methods still have their own limitations even for single-behavioral sequences. Markov Chain (MC) based models \cite{yang2010personalizing}\cite{rendle2010factorizing}\cite{natarajan2013app} have become the most popular methods for sequential prediction. MC based models aim to predict the users' next behavior based on the past behaviors. A transition matrix is estimated, which can give the probability of an action based on the previous ones. However, a major problem of MC based models is that all the components are independently combined, indicating that it makes strong independence assumption among multiple factors \cite{wang2015learning}.

Recently, Recurrent Neural Networks (RNN) have been successfully employed to model temporal dependency for different applications, such as sentence modeling tasks \cite{mikolov2010recurrent}\cite{mikolov2011extensions}\cite{mikolov2011rnnlm}, video modeling \cite{du2015hierarchical}, sequential click prediction \cite{zhang2014sequential} and location prediction \cite{liu2016strnn}. When modeling the sequential data, RNN assumes that the temporal dependency changes monotonously along with positions in a sequence. This means that, one element, e.g., a word, a frame and a product, in a sequence usually has more significant effect than the previous one for prediction. Such rules may well model words in a sentence or frames in a video, since adjacent words or frames have significant correlation. The larger the distance between two words or two frames, the smaller the correlation. However, for behavior prediction tasks, this assumption does not confirm to complex real situations, especially for the most recent elements in historical sequences. Sometimes, several most recent elements have similar effects on users' next behavior. For instance, if you went to a gym, a restaurant and a shopping market yesterday morning, afternoon and evening respectively, these three behaviors may have similar effects on your behaviors today. Sometimes, most recent elements have more complex effects on the future. For instance, going to a gym yesterday has dominant effects on how you exercise today, and what you ate at a restaurant yesterday or what you bought at a shopping market yesterday can affect what you want to eat today a lot. There is no guarantee that one element has more or less significant effect than the previous one. The effects of most recent elements in modeling human behaviors are much more complicated than that in modeling sentences or videos. But RNN can only tell us that behaviors in yesterday morning have more significant effects than behaviors in yesterday afternoon, and behaviors in yesterday afternoon have more significant effects than behaviors in yesterday evening. Accordingly, we can say that, RNN cannot well model short-term contexts in a sequence.

Different from the recurrent architecture in RNN based language models \cite{mikolov2010recurrent}\cite{mikolov2011extensions}\cite{mikolov2011rnnlm}, the Log-BiLinear (LBL) model \cite{mnih2007three} represents each word in a sentence, i.e., each position in a sequence, with a specific matrix. It can better model the complex situations of local contexts in sequences. But when the sequence is too long, a maximal length is usually set. And in real behavior prediction scenarios, length of behavioral sequences is usually not fixed. So, LBL cannot well model long-term contexts in a sequence.

Furthermore, time difference between input elements, e.g., continuous time difference between apps that the user has behaviors on in Figure \ref{fig:Example}, is another key factor in sequential modeling. However, to our best knowledge, none of existing models, including above MC based methods, RNN and LBL, can jointly model sequential information and time difference information in one framework.

In this paper, to overcome above shortcomings of conventional methods and model multi-behavioral sequences, we propose two novel sequential prediction methods, i.e., \textbf{Recurrent Log-BiLinear (RLBL)} model and \textbf{Time-Aware Recurrent Log-BiLinear (TA-RLBL)} model. \textit{First}, to capture the properties of different types of behaviors in historical sequences, we employ behavior-specific transition matrices in our model. To the best of our knowledge, this is the first work which is designed for predicting multi-behavioral sequences \textit{Second}, we design RLBL model as a recurrent architecture to capture long-term contexts in sequences. It models several elements in each hidden layer and uses position-specific transition matrices to capture short-term contexts of the historical sequence. Our RLBL not only can model the subtle characteristics of the most recent items in a sequence, but also can deal with long-term contexts with a recurrent structure. \textit{Third}, we further extend the RLBL model based on time difference information, and propose the TA-RLBL model. Rather than specific matrices for each position in RLBL, we use specific matrices, i.e., time-specific transition matrices, for each time difference value between input elements in TA-RLBL. Since it is difficult to estimate matrices for all the continuous time difference values, we divide all the possible temporal values into discrete bins. For a specific time difference value in one time bin, we can calculate the corresponding transition matrix via a linear interpolation of transition matrices of the upper bound and lower bound. Incorporating continuous time difference information, TA-RLBL can further improve the performance of RLBL.

The main contributions of this work are listed as follows:
\begin{itemize}
\item
We firstly address the problem of multi-behavioral sequential prediction, which is a significant problem in sequential prediction. And we use behavior-specific matrices to represent the effects of different types of behaviors.

\item
The RLBL model incorporates position-specific matrices and the recurrent structure, which can well model both short- and long-term contexts in historical sequences.

\item
TA-RLBL uses time-specific matrices to jointly model sequential information and time difference information in one framework, which further improves the performance of RLBL.

\item
Experiments conducted on three real-world datasets show that RLBL and TA-RLBL are effective and clearly outperform state-of-the-art methods.

\end{itemize}

The rest of the paper is organized as follows. In section 2, we review some related work on sequential prediction. Then we give the problem definition of multi-behavioral sequential prediction in section 3. Section 4 and 5 detail our RLBL model and TA-RLBL model respectively. In section 6, we introduce the learning methods of our proposed models. In section 7, we conduct experiments on three real-world datasets and compare with several state-of-the-art methods. Section 8 concludes our work and discusses future research.

\section{Related Works}
In this section, we review several types of methods for sequential prediction and time-aware prediction, i.e., time-aware neighborhood based methods, time-aware factorization methods, markov chain based methods and neural network based methods.

\subsection{Time-aware Neighborhood}

Time-aware neighborhood models \cite{ding2005time}\cite{lathia2009temporal}\cite{liu2010online} may be the most natural methods for modeling sequential information. These methods employ neighborhood based algorithms to capture temporal effects via giving more relevance to recent observations and less to past observations. However, though these methods may confirm to our first instinct and properties of sequential information, neighborhood based methods are unable to reveal the underlying properties in users' historical sequences.

\subsection{Time-aware Factorization Methods}

Matrix factorization (MF) based methods \cite{mnih2007probabilistic}\cite{koren2009matrix}\cite{koren2011advances} have become the state-of-the-art approach to collaborative filtering. Nowadays, MF based methods have been extended for more general and complex situations \cite{rendle2012factorization}\cite{rendle2011fast}. Among them, time-aware factorization based models have been extensively studied. Tensor Factorization (TF) \cite{bahadori2014fast}\cite{xiong2010temporal} treats time slices as another dimension and generates latent vectors of time slices via factorization to capture the underlying properties in the behavioral history. TimeSVD++ \cite{koren2010collaborative} learns time-aware representations for users and items in different time slices. However, factorization based models have difficulties in generating latent representations for time slices which has never or seldom appeared in the training data. Thus, factorization based models are not able to accurately predict item in the future time slices.

\subsection{Markov Chain Based Methods}

Based on the Markov assumption, MC based methods are widely used models for sequential prediction tasks \cite{yang2010personalizing}. MC based models predict users' next behaviors via estimating a transition matrix, which gives the probability of an action based on the previous ones. Via factorization of the personalized probability transition matrices of users, Factorizing Personalized Markov Chain (FPMC) \cite{rendle2010factorizing} can provide more accurate prediction for each sequence. FPMC is also extended by using the user group \cite{natarajan2013app} or incorporating the location constraint \cite{cheng2013you}. Recently, some factors of human brain have been added into MC based methods, including interest-forgetting curve \cite{chen2015personalized} and dynamics of boredom \cite{kapoor2015just}. However, the main drawback of MC based models is the independent combination of the past components, which lies in a strong independence assumption and confines the prediction accuracy. Then MC based methods are extended by using representation learning. Hierarchical Representation Model (HRM) \cite{wang2015learning} learns the hierarchical representation of behaviors in the last transaction and in the past history of a user to predict behaviors in the next transaction. And Personalized Ranking Metric Embedding (PRME) \cite{feng2015personalized} learns embeddings of users according to distances between locations. These methods still face a problem that they only model items in the most recent history and previous items can only be modeled by constant user latent vectors. Thus, except items in the most recent history, other items after model training will be ignored. User representations cannot change dynamically along with behavioral sequences.

\subsection{Neural Network Based Methods}

Recently, some prediction models, especially language models \cite{mikolov2013distributed}, are proposed based on neural networks. The most classical neural language model is proposed via a single layer neural network \cite{bengio2003neural}. Among variety language models, RNN has been the most successful one in modeling sentences \cite{mikolov2010recurrent}\cite{mikolov2011extensions}\cite{mikolov2011rnnlm}. It has successfully applied in variety natural language processing tasks, such as machine translation \cite{cho2014learning}\cite{sutskever2014sequence}, conversation machine \cite{serban2016building}\cite{shang2015neural} and image caption \cite{mao2014deep}\cite{vinyals2015show}. Recently, RNN based models also achieve successive results in other areas. For video analyzing, RNN brings satisfying results for action recognition \cite{du2015hierarchical}. Incorporating users' each clicking as an input element of each layer, RNN has greatly improved the performance of sequential click prediction \cite{zhang2014sequential}. Spatial-Temporal Recurrent Neural Netwrks (ST-RNN) \cite{liu2016strnn} learns geographical distance-specific transition matrices in RNN framework for location prediction. And Dynamic REcurrent bAsket Model (DREAM) \cite{yu2016dream} uses pooling methods in each layer of RNN for aggregating items in one transaction and achieves state-of-the-art performance in next basket recommendation \cite{yu2016dream}. Context-Aware Recurrent Neural Netwrks (CA-RNN) \cite{liu2016context} incorporates variety of contextual information in the RNN structure for recommender systems. However, when modeling sequential data, RNN assumes that temporal dependency changes monotonously along with the positions in a sequence, which means one element in a sequence usually has more significant effect than the previous one for prediction. This is usually suitable for words in sentences or frames in videos. But it does not confirm to practical situations for predicting behaviors, especially for the most recent elements of a historical sequence. Several most recent elements may usually have similar or even more complex effects on a user's next choice. But RNN can only tell us that the most recent item has more significant effect than the previous items. So, we can say that RNN cannot well model short-term contexts in behavior modeling.

LBL \cite{mnih2007three} is another widely-used language model, which represents elements at each position in a sequence with specific matrices. And a hierarchical softmax \cite{mnih2009scalable} is utilized to accelerate LBL model. However, when sequences are too long, a maximal length is usually set and long-term contexts are discarded. So, LBL cannot well model long-term contexts in sequences, which often exist in real behavior prediction situations.

There also exist some studies on RNN based methods taking insight in modeling short-term and long-term contexts, e.g., Multi-timescale RNN \cite{yamashita2008emergence}\cite{liu2015multi} and Clockwork RNN \cite{koutnik2014clockwork}. Based on a hierarchical RNN structure \cite{el1995hierarchical}, these methods model short-term dependencies and long-term dependencies separately with multiple RNNs. These multiple RNNs are at different timescales, where the fastest one operates every input element, and relatively slower ones take delays and skip some input elements according to corresponding timescales. However, these RNN structures aim to better capture long-term dependencies in sequences via incorporating larger timescales in some of the many RNNs. Although they can indeed achieve better performance comparing with conventional structures in some applications \cite{yamashita2008emergence}\cite{koutnik2014clockwork}\cite{liu2015multi}, they still model input elements according to sequential orders in a RNN structure. Accordingly, they cannot overcome the drawback of RNN that temporal dependency changes monotonously. It is still hard for these methods to well model short-term contexts in behavior modeling scenarios.

\section{Problem Definition}

The multi-behavioral sequential prediction problem we study in this work can be formulated as follows. We have a set of users and a set of items denoted as $U  = \{ u_1 ,u_2 ,...\}$ and $V  = \{ v_1 ,v_2 ,...\}$ respectively. Multiple types of behaviors are denoted as $B = \{ b_1 ,b_2 ,...\}$. Each behavior of user $u$ is associated with a behavioral type and a timestamp. Then the sequential behavioral history of user $u$ consists of items $V^u = \{v^u_{1}, v^u_{2}, ...\}$, corresponding behavioral types $B^u = \{b^u_{1}, b^u_{2}, ...\}$ and timestamps $T^u = \{t^u_{1}, t^u_{2}, ...\}$. Given behavioral history of users towards items, the task is to predict what a specific user will choose next under a specific behavior.

\begin{figure}[tb]
\centering
\includegraphics[width=0.5\textwidth]{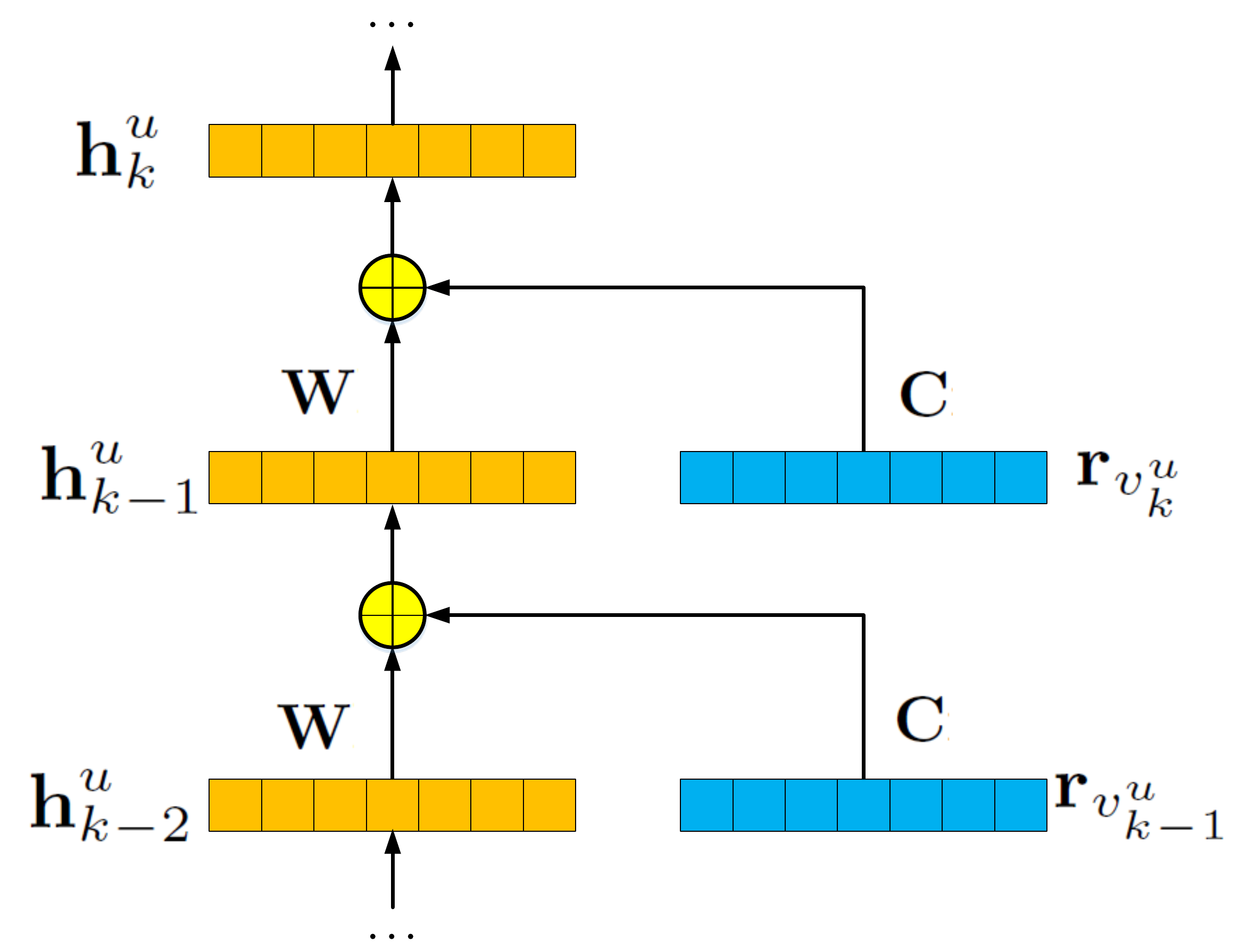}
\caption{Illustration of the Recurrent Neural Networks (RNN) model. RNN is a recurrent architecture with multiple hidden layers. The hidden status of RNN changes dynamically along with sequences, where the trend is monotonous. RNN has difficulty in modeling short-term contexts in behavioral sequences.}
\label{fig:RNN}
\end{figure}

\begin{figure}[tb]
\centering
\includegraphics[width=0.4\textwidth]{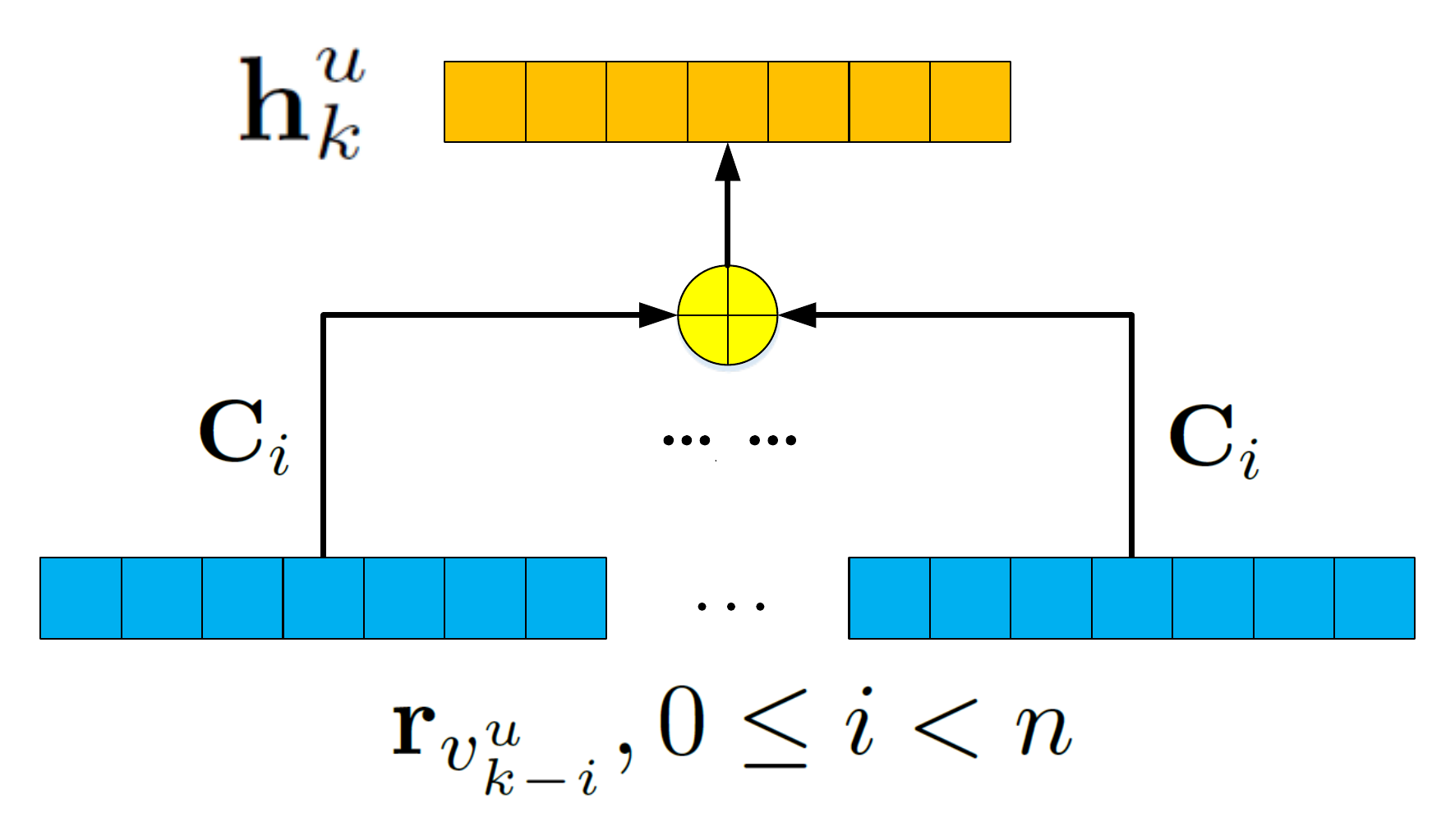}
\caption{Illustration of the Log-BiLinear (LBL) model. LBL is a feedforward neural network with a single linear hidden layer. In LBL, each position in sequences is modeled with a specific transition matrix. And a maximal number of modeled elements is usually set. LBL has difficulty in modeling long-term contexts in behavioral sequences.}
\label{fig:LBL}
\end{figure}

Here, taking the application in e-commerce as an example, there will be four types of behaviors (i.e., clicking, purchasing, adding to favorites and adding to shopping chart) denoted as $\{b_1 ,b_2 ,b_3 ,b_4\}$. The task is to predict which item a user would like to click, purchase, add to favorites or add to shopping chart next. Similarly, in the app usage, there will be three types of behaviors (i.e., downloading, using and uninstalling) denoted as $\{b_1 ,b_2 ,b_3\}$. Then the task becomes predicting which app a user would like to download, use or uninstall next.

\section{Recurrent Log-bilinear Model (RLBL)}

In this section, we present the recurrent log-bilinear model. We first introduce the RNN model and LBL model, then detail the architecture of RLBL with a single type of behaviors and introduce how RLBL can be employed to model multiple types of behaviors.

\subsection{Recurrent Neural Networks}

The architecture of RNN is shown in Figure \ref{fig:RNN}. It consists of an input layer, an output unit, multiple hidden layers, as well as inner weight matrices \cite{zhang2014sequential}. The activation values of the hidden layers are computed as:
\begin{equation}
\mathbf{h}^u_{k} = f\left( {\mathbf{W} \mathbf{h}^u_{k-1} + \mathbf{C} \mathbf{r}_{v^u_k} } \right)~,
\end{equation}
where $\mathbf{h}^u_{k} \in {\mathbb{R}^{d}}$ denotes the hidden representation of user $u$ at position $k$ in a sequence, $\mathbf{r}_{v^u_k} \in {\mathbb{R}^{d}}$ denotes the representation of the $k$th input item of user $u$. $f(x)$ is the activation function. $\mathbf{C} \in {\mathbb{R}^{d \times d}}$ and $\mathbf{W} \in {\mathbb{R}^{d \times d}}$ mean the transition matrix for the current items and the previous status respectively. $\mathbf{W}$ can propagate sequential signals, and $\mathbf{C}$ can capture users' current behavior. This activation process can be repeated iteratively and then the status at each position in a sequence can be calculated.

\subsection{Log-bilinear Model}

The Log-BiLinear (LBL) model \cite{mnih2007three} is a deterministic model that may be viewed as a feedforward neural network with a single linear hidden layer \cite{kiros2014multiplicative}. Using LBL for the sequential prediction problem, the final predicted representation of a sequence is generated based on the input items and the transition matrices at each position. As shown in Figure \ref{fig:LBL}, in the LBL model, the representation at next position is a linear prediction:
\begin{equation}
\mathbf{h}^u_{k} = \sum\limits_{i = 0}^{n - 1} {\mathbf{C}_i \mathbf{r}_{v^u_{k-i}} } ~,
\end{equation}
where $\mathbf{C}_i \in {\mathbb{R}^{d \times d}}$ denotes the transition matrix for the corresponding position in a sequence, and $n$ is the number of elements modeled in a sequence.

\begin{figure*}[!tb]
\centering
\subfigure[The RLBL model.]{
\begin{minipage}[b]{0.47\textwidth}
\includegraphics[width=1\textwidth]{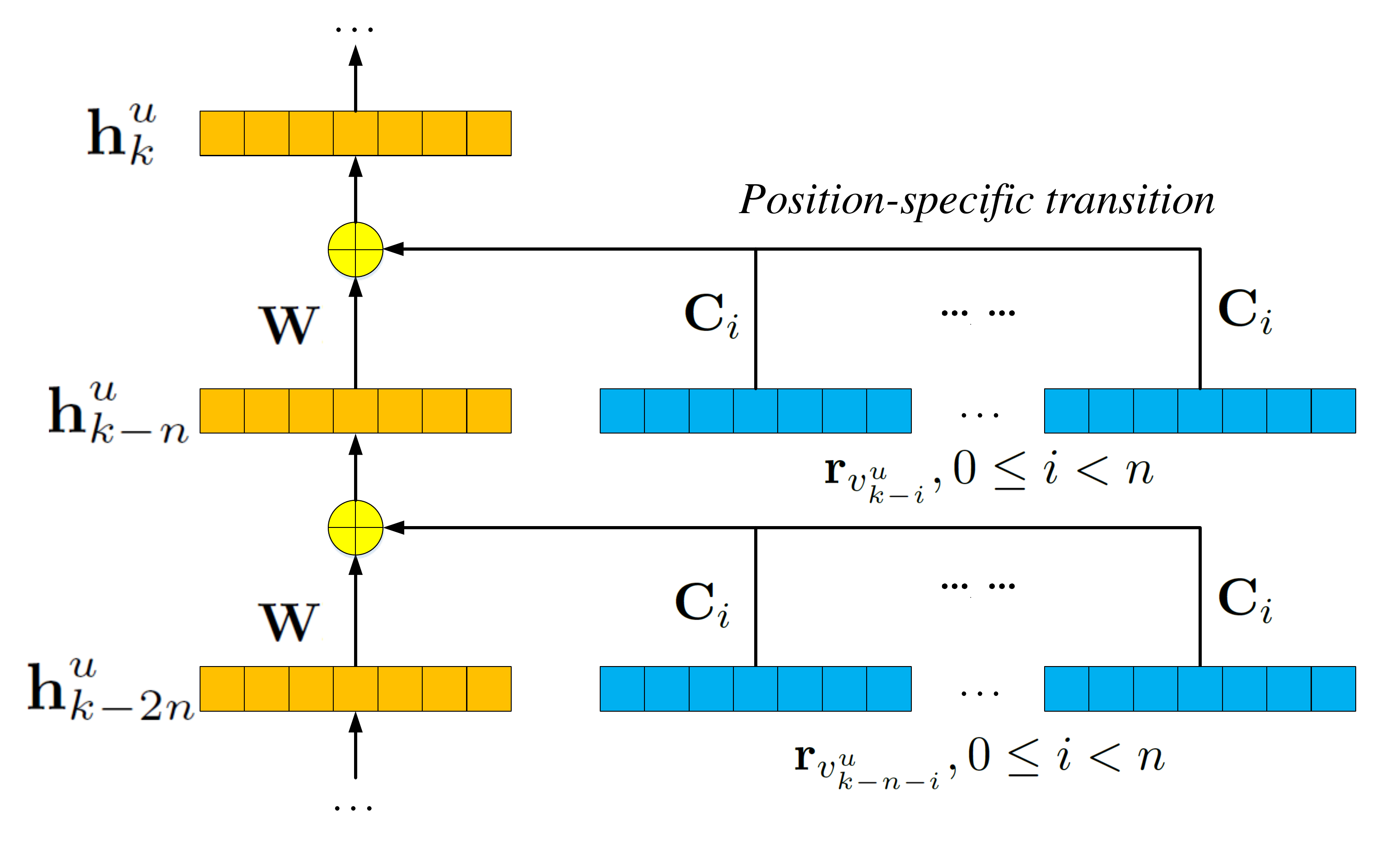}
\label{fig:RLBL}
\end{minipage}
}
\subfigure[The TA-RLBL model.]{
\begin{minipage}[b]{0.5\textwidth}
\includegraphics[width=1\textwidth]{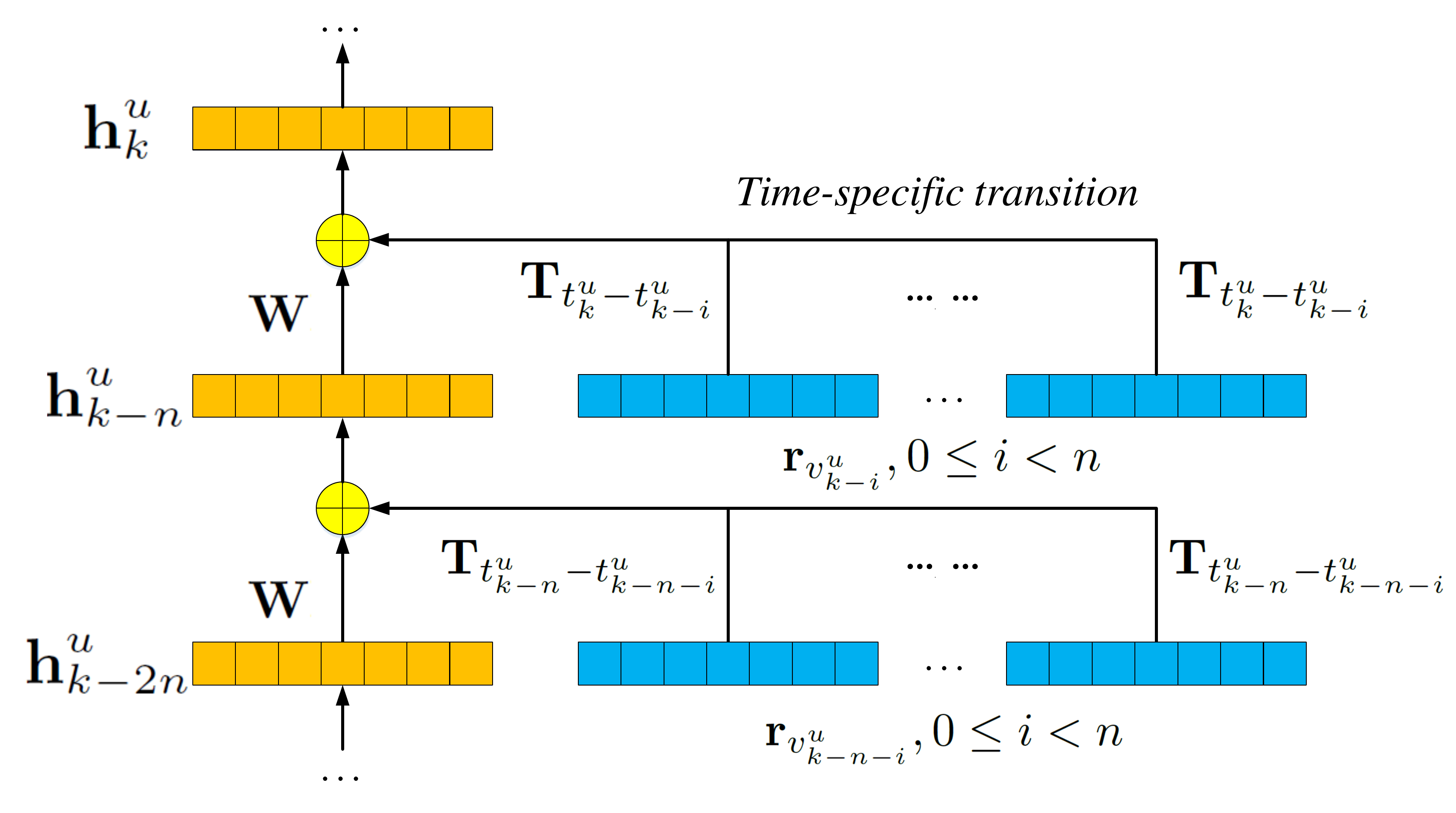}
\label{fig:TA-RLBL}
\end{minipage}
}
\caption{Illustration of the Recurrent Log-BiLinear (RLBL) model and the Time-Aware Recurrent Log-BiLinear (TA-RLBL) model. RLBL employs a recurrent architecture to capture long-term contexts. It models several elements in each hidden layer and incorporates position-specific transition matrices to capture short-term contexts in a historical sequence. TA-RLBL further extends the RLBL model. It replaces position-specific transition matrices with time-specific transition matrices to model time difference information. Behavior-specific matrices can be incorporated in RLBL and TA-RLBL to capture multiple types of behaviors in sequences.}
\label{fig:models}
\end{figure*}

\subsection{Modeling Single Type of Behaviors}

As discussed in the previous sections, though both RNN and LBL have achieved satisfying results, they still have their own drawbacks. RNN cannot well handle short-term contexts in a sequence, while LBL cannot well model long-term contexts.

To capture short-term and long-term contexts in historical sequences simultaneously, instead of modeling only one element in each hidden layer in RNN, we model several elements in each hidden layer and incorporate position-specific matrices into the recurrent architecture. As illustrated in Figure \ref{fig:RLBL}, given a user $u$, the hidden representation of the user at the position $k$ in a sequence can be computed as:
\begin{equation} \label{RLBL}
\mathbf{h}^u_{k}  = {\mathbf{W} \mathbf{h}^u_{k-n} + \sum\limits_{i = 0}^{n - 1} {\mathbf{C}_i \mathbf{r}_{v^u_{k-i}} } } ~,
\end{equation}
where $n$ is the number of input items modeled in one layer of RLBL, which is called the window width in this paper. The position-specific transition matrices $\mathbf{C}_i \in {\mathbb{R}^{d \times d}}$ captures the impact of short-term contexts, i.e., the $i$th item in one layer of RLBL, on user behaviors. And the characteristics of users' long-term history are modeled via the recurrent framework. Moreover, when we only consider one input item in each layer and set the window width $n=1$, the formulation of RLBL will be as the same as that of RNN ignoring the nonlinear activation function.

Notice that, when the sequence is shorter than the window width or the predicted position is at the very first part of a sequence, i.e., $k<n$. Equation \ref{RLBL} should be rewritten as:
\begin{equation}
\mathbf{h}^u_{k}  = {\mathbf{W} \mathbf{h}^u_{0} + \sum\limits_{i = 0}^{k - 1} {\mathbf{C}_i \mathbf{r}_{v^u_{k-i}} } } ~,
\end{equation}
where $\mathbf{h}^u_{0}  = \mathbf{u}_0$, denoting the initial status of users. The initial status of all users should be the same because personal information does not exist when a user has not selected an item. This representation $\mathbf{u}_0$ can be used to model cold start users. The equation in this special situation can be viewed as the same as that of a regular LBL model.

\subsection{Modeling Multiple Types of Behaviors}

Although there exist some scenarios with one type of behavior, e.g., purchasing in e-commerce and clicking on websites, there are much more applications with multiple types of behaviors towards items. For instance, users will click items, purchase items and add items to favorites in e-commerce. And users may download apps, use apps and uninstall apps. Thus, it is necessary to model multi-behavioral sequences and collaboratively predict what a user will choose next under a specific behavior.

We can simply ignore different types of behaviors, or treat different behaviors towards one item as different elements in conventional models. However, it is hard to model the correlation among different behaviors towards one item. Here, we incorporate behavior-specific matrices to capture properties of multiple types of behaviors. Then, the representation of user $u$ at position $k$ can be calculated as:
\begin{equation}
\mathbf{h}^u_{k}  = {\mathbf{W} \mathbf{h}^u_{k-n} + \sum\limits_{i = 0}^{n - 1} {\mathbf{C}_i \mathbf{M}_{b^u_{k-i}} \mathbf{r}_{v^u_{k-i}} } } ~,
\end{equation}
where $\mathbf{M}_{b^u_{i}} \in {\mathbb{R}^{d \times d}}$ denotes a behavior-specific transition matrix modeling the corresponding behavior on the $i$th item of user $u$. Note that, behavior-specific matrices can be omitted if there is only one type of behavior. Incorporating behavior-specific matrices, RLBL is the first approach which can be used to model the underlying properties of different types of behaviors in historical sequences.

Now, via calculating inner product, the prediction of whether user $u$ would conduct behavior $b$ on item $v$ at the sequential position $k+1$ can be made as:
\begin{equation} \label{prediction}
y_{u,k+1,b,v}  = (\mathbf{s}^u_k)^T \mathbf{M}_b \mathbf{r}_v  = (\mathbf{h}^u_{k} + \mathbf{u}_u)^T \mathbf{M}_b \mathbf{r}_v  ~,
\end{equation}
where $\mathbf{s}^u_k$ denotes the representation for the status of user $u$ at the sequential position $k$, containing dynamic representation $\mathbf{h}^u_{k}$ and static latent representation $\mathbf{u}_u \in {\mathbb{R}^{d}}$.

\section{Time-aware RLBL Model (TA-RLBL)}

Sequential models often ignore the continuous time difference between input elements. The time difference information is important for prediction considering that shorter time differences usually have more significant impact on the future comparing with longer time differences. For instance, suppose there are two items, $v_a$ and $v_b$, in a user's purchasing history. The user bought item $v_a$ last night and item $v_b$ last month. It is probably that the user's choice about what to buy next is mainly influenced by item $v_a$. In contrast, if item $v_b$ is bought last mourning, it is probably that both item $v_a$ and $v_b$ have similar impact to the user's choice because of similar interests in a short period. Moreover, as the purchasing behavior of some items is periodical such as buying tooth paste every month, the effect of time difference becomes more significant in such situations.

Accordingly, in this section, we extend our RLBL model with time difference information and introduce the time-aware recurrent log-bilinear model.

\subsection{Proposed Model}

As discussed above, it will be reasonable if we incorporate time difference information in our RLBL model. Here, we replace position-specific transition matrices with time-specific transition matrices and propose a time-aware RLBL model. As shown in Figure \ref{fig:TA-RLBL}, given a user $u$, representation at position $k$ can be calculated as:
\begin{equation} \label{TA-RLBL}
\mathbf{h}^u_{k}  = {\mathbf{W} \mathbf{h}^u_{k-n} + \sum\limits_{i = 0}^{n - 1} {\mathbf{T}_{t^u_{k}-t^u_{k-i}} \mathbf{r}_{v^u_{k-i}} } } ~,
\end{equation}
where $t^u_{k}$ denotes the current timestamp, $t^u_{k-i}$ means the timestamp of each item in one layer of TA-RLBL, and $\mathbf{T}_{t^u_{k}-t^u_{k-i}} \in {\mathbb{R}^{d \times d}}$ denotes the time-specific transition matrix for the time difference ${t^u_{k}-t^u_{k-i}}$ between timestamp $t^u_{k-i}$ and $t^u_{k}$. The time-specific transition can capture the time-aware impacts of the most recent behavioral history.

Moreover, similar to RLBL, when $k<n$, Equation \ref{TA-RLBL} should be rewritten as:
\begin{equation} \label{TA-RLBL-0}
\mathbf{h}^u_{k}  = {\mathbf{W} \mathbf{h}^u_{0} + \sum\limits_{i = 0}^{k - 1} {\mathbf{T}_{t^u_{k}-t^u_{k-i}} \mathbf{r}_{v^u_{k-i}} } } ~,
\end{equation}
where $\mathbf{h}^u_{0}  = \mathbf{u}_0$, denoting the initial status of users. To model multiple types of behavior, behavior-specific transition matrices are also applied in TA-RLBL model:
\begin{equation} \label{M-TA-RLBL}
\mathbf{h}^u_{k}  = {\mathbf{W} \mathbf{h}^u_{k-n} + \sum\limits_{i = 0}^{n - 1} {\mathbf{T}_{t^u_{k}-t^u_{k-i}} \mathbf{M}_{b^u_{k-i}} \mathbf{r}_{v^u_{k-i}} } } ~.
\end{equation}
Then, similar to RLBL, the prediction of whether user $u$ would conduct behavior $b$ on item $v$ at sequential position $k+1$ can be computed as:
\begin{equation} \label{prediction2}
y_{u,k+1,b,v}  = (\mathbf{s}^u_k)^T \mathbf{M}_b \mathbf{r}_v  = (\mathbf{h}^u_{k} + \mathbf{u}_u)^T \mathbf{M}_b \mathbf{r}_v  ~.
\end{equation}

\subsection{Linear Interpolation for Learning Transition Matrices}

If we learn a distinct matrix for each possible continuous time difference value, we have to estimate a great number of time-specific transition matrices and the model tends to overfit. Here, similar to the method in \cite{liu2016strnn}, we equally partition the range of all the possible time difference values into discrete bins. Only the transition matrices of the upper and lower bounds of time bins are needed to be estimated in our model. For time difference values in a time bin, their transition matrices can be calculated via a linear interpolation. Mathematically, the time-specific transition matrix $\mathbf{T}_{t_d}$ for time difference value $t_d$ can be calculated as:
\begin{equation}
\mathbf{T}_{t_d}  = \frac{{\left[ {\mathbf{T}_{L(t_d)} (U(t_d) - t_d) + \mathbf{T}_{U(t_d)} (t_d - L(t_d))} \right]}}{{\left[ {(U(t_d) - t_d) + (t_d - L(t_d))} \right]}} ~,
\end{equation}
where $U(t_d)$ and $L(t_d)$ denote the upper bound and lower bound of time difference $t_d$, $\mathbf{T}_{U(t_d)}$ and $\mathbf{T}_{L(t_d)}$ denote the time-specific transition matrices for $U(t_d)$ and $L(t_d)$ respectively. Such a linear interpolation method can solve the problem of learning time-specific transition matrices for continuous time differences. To be noted, although the change of time-specific matrices in each discrete time bin is linear, the global change in the entire range of all the possible time difference values is nonlinear.

For instance, if the range of all the possible time difference values is partitioned into one-hour bins, and we want to calculate the transition matrix for time difference value $1.6h$, the upper bound and lower bound of $1.6h$ will be $2h$ and $1h$ respectively, and the corresponding time-specific transition matrix $T_{1.6h}$ can be calculated as:
\begin{equation}
\begin{aligned}
&\mathbf{T}_{1.6h}  = \frac{{\left[ {\mathbf{T}_{1h} (2h - 1.6h) + \mathbf{T}_{2h} (1.6h - 1h)} \right]}}{{\left[ {(2h - 1.6h) + (1.6h - 1h)} \right]}} \\
&~~~~~~~~~= 0.4\mathbf{T}_{1h} + 0.6\mathbf{T}_{2h} \\
 \end{aligned}~.
\end{equation}

Until now, we have detailed the RLBL and TA-RLBL model. Both models can well capture sequential information. If there exists explicit time information, TA-RLBL model is more suitable than that of RLBL model. And if the dataset is not associated with detailed time information, RLBL mode will be more suitable than TA-RLBL model. Both models are constructed under the same framework and can be applied according to actual situations.

\section{Parameter Learning}

In this section, we introduce the learning process of our proposed RLBL and TA-RLBL model with Bayesian Personalized Ranking (BPR) \cite{rendle2009bpr} and Back Propagation Through Time (BPTT) \cite{rumelhart1988learning}.

\subsection{Learning of RLBL}

BPR \cite{rendle2009bpr} is a state-of-the-art pairwise ranking framework for the implicit feedback data. BPR has been used as objective function for learning of RNN based models in behavioral prediction tasks \cite{liu2016strnn}\cite{yu2016dream}. The basic assumption of BPR is that a user prefers a selected element than a negative one. Formally, we need to maximize the following probability:
\begin{equation} \label{Learning of RLBL}
p(u,k+1,b,v \succ v') = g(y_{u,k+1,b,v}  - y_{u,k+1,b,v'} ) ~,
\end{equation}
where $v'$ denotes a negative sample, and $g(x)$ is a nonlinear function which is selected as:
\begin{equation} \label{Learning of RLBL}
g(x) = \frac{1}{{1 + e^{ - x} }} ~.
\end{equation}
Incorporating the negative log likelihood, we can minimize the following objective function equivalently:
\begin{equation} \label{Objective}
J_1 = \sum {\ln (1 + e^{ - (y_{u,k+1,b,v}  - y_{u,k+1,b,v'} )} )}  + \frac{\lambda }{2}\left\| \mathbf{\Theta}_1  \right\|^2 ~,
\end{equation}
where $\mathbf{\Theta}_1  = \left\{ {\mathbf{U},\mathbf{R},\mathbf{W},\mathbf{C},\mathbf{M}} \right\}$ denotes all the parameters to be estimated, $\lambda $ is a parameter to control the power of regularization. And the derivations of $J_1$ with respect to the parameters can be calculated as:
\begin{displaymath}
\frac{{\partial J_1 }}{{\partial {\mathbf{u}}_u }} = \sum {\frac{{{\mathbf{M}}_b ({\mathbf{r}}_{v'}  - {\mathbf{r}}_v )l(u,k+1,b,v \succ v') }}{{1 + l(u,k+1,b,v \succ v') }}}  + \lambda {\mathbf{u}}_u ~,
\end{displaymath}
\begin{displaymath}
\frac{{\partial J_1 }}{{\partial {\mathbf{r}}_v }} =  - \sum {\frac{{\left( {{\mathbf{M}}_b } \right)^T ({\mathbf{h}}_k^u  + {\mathbf{u}}_u )l(u,k+1,b,v \succ v') }}{{1 + l(u,k+1,b,v \succ v') }}}  + \lambda {\mathbf{r}}_v  ~,
\end{displaymath}
\begin{displaymath}
\frac{{\partial J_1 }}{{\partial {\mathbf{r}}_{v'} }} = \sum {\frac{{\left( {{\mathbf{M}}_b } \right)^T ({\mathbf{h}}_k^u  + {\mathbf{u}}_u )l(u,k+1,b,v \succ v') }}{{1 + l(u,k+1,b,v \succ v') }}}  + \lambda {\mathbf{r}}_{v'}  ~,
\end{displaymath}
\begin{displaymath}
\frac{{\partial J_1 }}{{\partial {\mathbf{M}}_b }} = \sum {\frac{{({\mathbf{h}}_k^u  + {\mathbf{u}}_u )({\mathbf{r}}_{v'}  - {\mathbf{r}}_v )^T l(u,k+1,b,v \succ v') }}{{1 + l(u,k+1,b,v \succ v') }}}  + \lambda {\mathbf{M}}_b  ~,
\end{displaymath}
\begin{displaymath}
\frac{{\partial J_1 }}{{\partial {\mathbf{h}}_k^u }} =  - \sum {\frac{{{\mathbf{M}}_b ({\mathbf{r}}_{v'}  - {\mathbf{r}}_v )l(u,k+1,b,v \succ v') }}{{1 + l(u,k+1,b,v \succ v') }}}  ~,
\end{displaymath}
where
\begin{displaymath}
l(u,k+1,b,v \succ v') = e^{ - (y_{u,k + 1,b,v}  - y_{u,k + 1,b,v'} )} ~.
\end{displaymath}

The derivations of the output layer have been calculated. Under each layer of the recurrent structure, similar to the conventional RNN model, RLBL can be trained by using the Back Propagation Through Time (BPTT) algorithm \cite{rumelhart1988learning}, which has been used in practical sequential prediction models \cite{liu2016strnn}\cite{zhang2014sequential}. For user $u$, given the derivation $\frac{{\partial {J_1}}}{{\partial {\mathbf{h}^u_{k}}}}$ of the representation $\mathbf{h}^u_{k}$ at sequential position $k$, the corresponding gradient of parameters at the hidden layer can be calculated as:
\begin{displaymath}
\frac{{\partial J_1 }}{{\partial \mathbf{h}^u_{k-n} }} = \mathbf{W}^T  \frac{{\partial J_1 }}{{\partial \mathbf{h}^u_{k}}} ~,
\end{displaymath}
\begin{displaymath}
\frac{{\partial J_1 }}{{\partial \mathbf{W}^T}} = \frac{{\partial J_1 }}{{\partial \mathbf{h}^u_{k} }} (\mathbf{h}^u_{k-n})^T ~,
\end{displaymath}
\begin{displaymath}
\frac{{\partial J_1 }}{{\partial \mathbf{r}_{v^u_{k-i}} }} = (\mathbf{M}_{b^u_{k-i}})^T (\mathbf{C}_i)^T \frac{{\partial J_1 }}{{\partial \mathbf{h}^u_{k} }} ~,
\end{displaymath}
\begin{displaymath}
\frac{{\partial J_1 }}{{\partial \mathbf{C}_i }} = \frac{{\partial J_1 }}{{\partial \mathbf{h}^u_{k} }} (\mathbf{r}_{v^u_{k-i}})^T (\mathbf{M}_{b^u_{k-i}})^T ~,
\end{displaymath}
\begin{displaymath}
\frac{{\partial J_1 }}{{\partial \mathbf{M}_{b^u_{k-i}} }} = (\mathbf{C}_i)^T \frac{{\partial J_1 }}{{\partial \mathbf{h}^u_{k} }} (\mathbf{r}_{v^u_{k-i}})^T ~.
\end{displaymath}
This process can be repeated iteratively, and the gradients of all the parameters are obtained. Then, the model can be learned via Stochastic Gradient Descent (SGD) until converge.

\begin{table*}[tb]
  \centering \scriptsize
  \caption{Experimental summarization.}
    \begin{tabular}{ccccc}
    \toprule
    dataset & scenario & \#behavioral types & behaviors & behavior to predict \\
    \midrule
    Movielens & watching movies & 5     & rating 5, 4, 3, 2, 1 stars & rating 5 or 4 stars \\
    Global Terrorism Database   & terrorist attack & 7     & armed, unarmed, assassination, bombing, facility, hijacking, hostage & attack (all types) \\
    Tmall & e-commerce & 4     & clicking, purchasing, adding to favorites, adding to shopping cart & purchasing \\
    \bottomrule
    \end{tabular}%
  \label{tab:summarization}%
\end{table*}%

\subsection{Learning of TA-RLBL}

For learning of TA-RLBL, using BPR \cite{rendle2009bpr}, similar to Equation \ref{Learning of RLBL} and \ref{Objective}, we need to minimize the following objective function:
\begin{equation} \label{Objective2}
J_2 = \sum {\ln (1 + e^{ - (y_{u,k+1,b,v}  - y_{u,k+1,b,v'} )} )}  + \frac{\lambda }{2}\left\| \mathbf{\Theta}_2  \right\|^2 ~,
\end{equation}
where $\mathbf{\Theta_2}  = \left\{ {\mathbf{U},\mathbf{R},\mathbf{W},\mathbf{T},\mathbf{M}} \right\}$ denotes all the parameters to be estimated in TA-RLBL. Similarly, the derivations of $J_2$ with respect to the parameters can be computed as:
\begin{displaymath}
\frac{{\partial J_2 }}{{\partial {\mathbf{u}}_u }} = \sum {\frac{{{\mathbf{M}}_b ({\mathbf{r}}_{v'}  - {\mathbf{r}}_v )l(u,k+1,b,v \succ v') }}{{1 + l(u,k+1,b,v \succ v') }}}  + \lambda {\mathbf{u}}_u ~,
\end{displaymath}
\begin{displaymath}
\frac{{\partial J_2 }}{{\partial {\mathbf{r}}_v }} =  - \sum {\frac{{\left( {{\mathbf{M}}_b } \right)^T ({\mathbf{h}}_k^u  + {\mathbf{u}}_u )l(u,k+1,b,v \succ v') }}{{1 + l(u,k+1,b,v \succ v') }}}  + \lambda {\mathbf{r}}_v  ~,
\end{displaymath}
\begin{displaymath}
\frac{{\partial J_2 }}{{\partial {\mathbf{r}}_{v'} }} = \sum {\frac{{\left( {{\mathbf{M}}_b } \right)^T ({\mathbf{h}}_k^u  + {\mathbf{u}}_u )l(u,k+1,b,v \succ v') }}{{1 + l(u,k+1,b,v \succ v') }}}  + \lambda {\mathbf{r}}_{v'}  ~,
\end{displaymath}
\begin{displaymath}
\frac{{\partial J_2 }}{{\partial {\mathbf{M}}_b }} = \sum {\frac{{({\mathbf{h}}_k^u  + {\mathbf{u}}_u )({\mathbf{r}}_{v'}  - {\mathbf{r}}_v )^T l(u,k+1,b,v \succ v') }}{{1 + l(u,k+1,b,v \succ v') }}}  + \lambda {\mathbf{M}}_b  ~,
\end{displaymath}
\begin{displaymath}
\frac{{\partial J_2 }}{{\partial {\mathbf{h}}_k^u }} =  - \sum {\frac{{{\mathbf{M}}_b ({\mathbf{r}}_{v'}  - {\mathbf{r}}_v )l(u,k+1,b,v \succ v') }}{{1 + l(u,k+1,b,v \succ v') }}}  ~,
\end{displaymath}
where
\begin{displaymath}
l(u,k+1,b,v \succ v') = e^{ - (y_{u,k + 1,b,v}  - y_{u,k + 1,b,v'} )} ~.
\end{displaymath}

Then, similar to RLBL, using BPTT \cite{rumelhart1988learning}, for user $u$, given the derivation $\frac{{\partial {J_2}}}{{\partial {\mathbf{h}^u_{k}}}}$ of the representation $\mathbf{h}^u_{k}$ at the sequential position $k$, the corresponding gradient of parameters at the hidden layer can be calculated as:
\begin{displaymath}
\frac{{\partial J_2 }}{{\partial \mathbf{h}^u_{k-n} }} = \mathbf{W}^T  \frac{{\partial J_2 }}{{\partial \mathbf{h}^u_{k}}} ~,
\end{displaymath}
\begin{displaymath}
\frac{{\partial J_2 }}{{\partial \mathbf{W}^T}} = \frac{{\partial J_2 }}{{\partial \mathbf{h}^u_{k} }} (\mathbf{h}^u_{k-n})^T ~,
\end{displaymath}
\begin{displaymath}
\frac{{\partial J_2 }}{{\partial \mathbf{r}_{v^u_{k-i}} }} = (\mathbf{M}_{b^u_{k-i}})^T (\mathbf{T}_{t^u_{k}-t^u_{k-i}})^T \frac{{\partial J_2 }}{{\partial \mathbf{h}^u_{k} }} ~,
\end{displaymath}
\begin{displaymath}
\frac{{\partial J_2 }}{{\partial \mathbf{T}_{t^u_{k}-t^u_{k-i}} }} = \frac{{\partial J_2 }}{{\partial \mathbf{h}^u_{k} }} (\mathbf{r}_{v^u_{k-i}})^T (\mathbf{M}_{b^u_{k-i}})^T ~,
\end{displaymath}
\begin{displaymath}
\frac{{\partial J_2 }}{{\partial \mathbf{M}_{b^u_{k-i}} }} = (\mathbf{T}_{t^u_{k}-t^u_{k-i}})^T \frac{{\partial J_2 }}{{\partial \mathbf{h}^u_{k} }} (\mathbf{r}_{v^u_{k-i}})^T ~.
\end{displaymath}
The process above can be repeated iteratively, and we can obtain all the gradients. After that, the model can be trained via SGD until converge.

\section{Experiments}

In this section, we empirically investigate the performance of RLBL and TA-RLBL. As shown in Table \ref{tab:summarization}, we conduct our experiments on three scenarios with different numbers of behavioral types. We first introduce our experimental settings. Then we conduct experiments to compare RLBL and TA-RLBL with different window width and experiments to compare performances of single behavior and multiple behaviors. We also give comparison of our models and some state-of-the-art methods with varying dimensionality. Then, we study the performance of models under different length of behavioral history. Finally, we analyse the computational time and convergence of our proposed methods.

\subsection{Experimental Settings}
Our experiments are conducted on three real datasets with different numbers of behavioral types. Details of these datasets are illustrated in Table \ref{tab:summarization}.
\begin{itemize}
\item \textbf{Movielens-1M}\footnote{http://grouplens.org/datasets/movielens/} is a widely used dataset, associated with timestamps, for the rating prediction in recommender systems. It contains about 1,000,000 rating records of 4,000 movies by 6,000 users. The ratings are divided into five levels, indicating users' different levels of preference, which can be viewed as five different types of behaviors. With this dataset, we aim to predict which movie a user will rate 5 or 4 stars next, i.e., which movie a user will prefer next.

\item \textbf{Global Terrorism Database}\footnote{http://www.start.umd.edu/gtd/} includes more than 125,000 terrorist incidents that have occurred all around the world since 1970 conducted by about 3,000 terrorist organizations. This dataset consists of 7 behavioral types, i.e., different attacking types, as indicated in Table \ref{tab:summarization}. For social good, we would like to predict which province or state a terrorist organization will attack. Thus, it is available for us to take action before accidents happen and save people's life.

\item \textbf{Tmall}\footnote{https://102.alibaba.com/competition/addDiscovery/index.htm} is a dataset collected from Tmall\footnote{https://www.tmall.com/}, one of the biggest online shopping websites in China. It contains about 200,000 shopping records belonging to 1,000 users on 10,000 items. The temporal information of the dataset is extracted based on the day level. It contains four different types of behaviors: clicking, purchasing, adding to favorites and adding to shopping cart. It suits for the task of collaborative prediction on multi-behavioral sequences. On this dataset, we aim to predict what users will purchase next.
\end{itemize}

For each behavioral sequence of these three datasets, we use first $70\%$ of the items in the sequence for training, following $10\%$ data as the validation set for tuning parameters, e.g., the dimensionality of latent representations, and remaining $20\%$ for testing. The regularization parameter is set as $\lambda = 0.01$. And we use line search to select learning rates in each iteration.

\begin{table*}[!tb]
\caption{Comparison of RLBL and TA-RLBL with varying window width $n$ and dimensionality $d=8$.}
\centering
\subtable[Performance on the Movielens dataset.]{
    \begin{tabular}{ccccccccccc}
    \toprule
    method & $n$     & recall@1 & recall@2 & recall@5 & recall@10 & F1-score@1 & F1-score@2 & F1-score@5 & F1-score@10 & MAP \\
    \midrule
    \multirow{7}[0]{*}{RLBL} & 2     & 0.0067  & 0.0103  & 0.0333  & 0.0508  & 0.0067  & 0.0069  & 0.0111  & 0.0093  & 0.0377  \\
          & 3     & 0.0070  & 0.0104  & 0.0334  & 0.0510  & 0.0070  & 0.0070  & 0.0111  & 0.0093  & 0.0381  \\
          & 4     & 0.0070  & 0.0107  & 0.0338  & 0.0520  & 0.0070  & 0.0072  & 0.0113  & 0.0095  & 0.0385  \\
          & 5     & 0.0070  & 0.0108  & 0.0343  & 0.0527  & 0.0070  & 0.0072  & 0.0114  & 0.0096  & 0.0386  \\
          & 6     & 0.0071  & 0.0112  & 0.0354  & 0.0538  & 0.0071  & 0.0074  & 0.0118  & 0.0098  & 0.0395  \\
          & 7     & 0.0070  & 0.0111  & 0.0354  & 0.0543  & 0.0070  & 0.0074  & 0.0118  & 0.0099  & 0.0393  \\
          & 8     & 0.0070  & 0.0108  & 0.0351  & 0.0535  & 0.0070  & 0.0072  & 0.0117  & 0.0097  & 0.0390  \\
    \midrule
    \multirow{7}[0]{*}{TA-RLBL} & 2     & 0.0070  & 0.0106  & 0.0343  & 0.0529  & 0.0070  & 0.0071  & 0.0114  & 0.0096  & 0.0388  \\
          & 3     & 0.0071  & 0.0105  & 0.0338  & 0.0523  & 0.0071  & 0.0071  & 0.0113  & 0.0094  & 0.0385  \\
          & 4     & 0.0071  & 0.0108  & 0.0337  & 0.0522  & 0.0071  & 0.0073  & 0.0113  & 0.0095  & 0.0388  \\
          & 5     & 0.0070  & 0.0110  & 0.0366  & 0.0553  & 0.0068  & 0.0074  & 0.0123  & 0.0101  & 0.0396  \\
          & 6     & \textbf{0.0072} & \textbf{0.0115} & \textbf{0.0372} & \textbf{0.0554} & \textbf{0.0072} & \textbf{0.0076} & \textbf{0.0124} & \textbf{0.0101} & \textbf{0.0404} \\
          & 7     & 0.0070  & 0.0115  & 0.0362  & 0.0549  & 0.0070  & 0.0076  & 0.0121  & 0.0100  & 0.0398  \\
          & 8     & 0.0070  & 0.0110  & 0.0348  & 0.0539  & 0.0070  & 0.0073  & 0.0118  & 0.0100  & 0.0392  \\
    \bottomrule
    \end{tabular}%
  \label{tab:mine_movielens}%
}
\qquad
\subtable[Performance on the Global Terrorism Database.]{
    \begin{tabular}{ccccccccccc}
    \toprule
    method & $n$     & recall@1 & recall@2 & recall@5 & recall@10 & F1-score@1 & F1-score@2 & F1-score@5 & F1-score@10 & MAP \\
    \midrule
    \multirow{7}[0]{*}{RLBL} & 2     & 0.1577  & 0.2448  & 0.4378  & 0.6104  & 0.1577  & 0.1632  & 0.1459  & 0.1110  & 0.2930  \\
          & 4     & 0.1642  & 0.2691  & 0.4676  & 0.6395  & 0.1642  & 0.1794  & 0.1559  & 0.1163  & 0.3082  \\
          & 6     & 0.1624  & 0.2686  & 0.4768  & 0.6468  & 0.1624  & 0.1791  & 0.1589  & 0.1176  & 0.3090  \\
          & 9     & 0.1580  & 0.2848  & 0.4865  & \textbf{0.6748} & 0.1580  & 0.1899  & 0.1622  & \textbf{0.1227} & 0.3153  \\
          & 10    & 0.1569  & 0.2806  & 0.4846  & 0.6659  & 0.1569  & 0.1871  & 0.1615  & 0.1211  & 0.3130  \\
          & 15    & 0.1567  & 0.2660  & 0.4682  & 0.6470  & 0.1567  & 0.1773  & 0.1561  & 0.1176  & 0.3053  \\
          & 20    & 0.1690  & 0.2775  & 0.4872  & 0.6572  & 0.1690  & 0.1850  & 0.1624  & 0.1195  & 0.3165  \\
    \midrule
    \multirow{7}[0]{*}{TA-RLBL} & 2     & 0.1642  & 0.2763  & 0.4740  & 0.6451  & 0.1697  & 0.1842  & 0.1580  & 0.1173  & 0.3117  \\
          & 4     & 0.1681  & 0.2758  & 0.4719  & 0.6411  & 0.1686  & 0.1839  & 0.1573  & 0.1166  & 0.3187  \\
          & 6     & 0.1678  & 0.2758  & 0.4833  & 0.6524  & 0.1678  & 0.1839  & 0.1611  & 0.1186  & 0.3146  \\
          & 9     & 0.1634  & \textbf{0.2895} & \textbf{0.4926} & 0.6730  & 0.1634  & \textbf{0.1930} & \textbf{0.1642} & 0.1224  & \textbf{0.3199} \\
          & 10    & 0.1622  & 0.2864  & 0.4910  & 0.6672  & 0.1622  & 0.1909  & 0.1637  & 0.1213  & 0.3180  \\
          & 15    & 0.1618  & 0.2731  & 0.4746  & 0.6527  & 0.1618  & 0.1821  & 0.1582  & 0.1187  & 0.3107  \\
          & 20    & \textbf{0.1697} & 0.2849  & 0.4839  & 0.6629  & \textbf{0.1697} & 0.1899  & 0.1613  & 0.1205  & 0.3197  \\
    \bottomrule
    \end{tabular}%
  \label{tab:mine_GTD}%
}
\qquad
\subtable[Performance on the Tmall dataset.]{
    \begin{tabular}{ccccccccccc}
    \toprule
    method & $n$     & recall@1 & recall@2 & recall@5 & recall@10 & F1-score@1 & F1-score@2 & F1-score@5 & F1-score@10 & MAP \\
    \midrule
    \multirow{7}[0]{*}{RLBL} & 2     & 0.1507  & 0.2170  & 0.3712  & 0.4690  & 0.1507  & 0.1447  & 0.1237  & 0.0853  & 0.2704  \\
          & 3     & 0.1480  & 0.2515  & \textbf{0.4118} & 0.5176  & 0.1480  & 0.1677  & \textbf{0.1373} & 0.0941  & 0.2781  \\
          & 4     & 0.1467  & 0.2311  & 0.3646  & 0.4953  & 0.1467  & 0.1541  & 0.1215  & 0.0901  & 0.2689  \\
          & 5     & \textbf{0.1600} & 0.2158  & 0.3975  & 0.5519  & \textbf{0.1600} & 0.1439  & 0.1325  & 0.1003  & \textbf{0.2836} \\
          & 6     & 0.1502  & 0.2272  & 0.3822  & \textbf{0.5596} & 0.1502  & 0.1515  & 0.1274  & \textbf{0.1017} & 0.2806  \\
          & 7     & 0.1493  & \textbf{0.2553} & 0.4074  & 0.5272  & 0.1493  & \textbf{0.1702} & 0.1358  & 0.0959  & 0.2819  \\
          & 8     & 0.1387  & 0.2324  & 0.4019  & 0.5395  & 0.1387  & 0.1549  & 0.1340  & 0.0981  & 0.2770  \\
    \midrule
    \multirow{7}[0]{*}{TA-RLBL} & 2     & 0.1351  & 0.2302  & 0.3669  & 0.4493  & 0.1351  & 0.1535  & 0.1223  & 0.0817  & 0.2608  \\
          & 3     & 0.1268  & 0.1931  & 0.3497  & 0.4541  & 0.1268  & 0.1287  & 0.1166  & 0.0826  & 0.2579  \\
          & 4     & 0.1441  & 0.2450  & 0.4084  & 0.4780  & 0.1441  & 0.1633  & 0.1361  & 0.0869  & 0.2820  \\
          & 5     & 0.1413  & 0.2366  & 0.3871  & 0.5461  & 0.1413  & 0.1577  & 0.1290  & 0.0993  & 0.2804  \\
          & 6     & 0.1253  & 0.2039  & 0.4081  & 0.4454  & 0.1253  & 0.1359  & 0.1360  & 0.0810  & 0.2521  \\
          & 7     & 0.1234  & 0.2213  & 0.3556  & 0.4412  & 0.1234  & 0.1475  & 0.1185  & 0.0802  & 0.2523  \\
          & 8     & 0.1198  & 0.2063  & 0.3472  & 0.4247  & 0.1198  & 0.1375  & 0.1157  & 0.0772  & 0.2454  \\
    \bottomrule
    \end{tabular}%
  \label{tab:mine_tmall}%
}
\label{tab:mine}%
\end{table*}

We compare RLBL and TA-RLBL with both conventional and state-of-the-art sequential methods.
\begin{itemize}
\item \textbf{POP} is a naive baseline method that recommends the most popular items to users.
\item \textbf{MF} \cite{mnih2007probabilistic} is one of the state-of-the-art methods for conventional collaborative filtering. It factorizes a user-item matrix into two low rank matrices, each of which represents the latent factors of users or items.
\item \textbf{MC} is a classical sequential model based on markov assumption, and is used as a sequential baseline method.
\item \textbf{TF} \cite{xiong2010temporal} is an extension of MF method. It extends MF from two dimensions to three dimensions, and the temporal information is modeled as the additional dimension.
\item \textbf{FPMC} \cite{rendle2010factorizing} extends conventional MC methods and factorizes personalized probability transition matrices of users. It is a widely-used method for sequential prediction and next basket recommendation.
\item \textbf{HRM} \cite{wang2015learning} learns the representation of behaviors in the previous transaction and predicts next behaviors. It has become a state-of-the-art method for next basket recommendation.
\item \textbf{RNN} \cite{yu2016dream} is a state-of-the-art method for the sequential prediction. It has been successfully applied in some applications, such as sentence modeling, click prediction, location prediction and next basket recommendation.
\end{itemize}

Considering TF learns latent vectors for time slices, and MC, FPMC and HRM predict future behaviors according to behaviors in the last transaction, we need to split transactions in different datasets according to corresponding application scenarios. So, we set the length of transaction in the Movielens dataset, the Global Terrorism Database and the Tmall dataset as one week, one month and one day respectively.

As above methods cannot model multi-behavioral sequences, when conducting compared methods on multi-behavioral datasets, we ignore different types of behaviors in behavioral histories. This means we treat different behaviors towards one item as the same.

Moreover, to investigate the performance of our proposed methods and compared methods, we select several widely-used evaluation metrics for our experiments.

\begin{itemize}
\item \textbf{Recall@k} and \textbf{F1-score@k} are two important metrics for ranking tasks. The evaluation score for our experiments is computed according to where the next selected item appears in the predicted list. We report recall@k and F1-score@k with $k=1$, $2$, $5$ and $10$ in our experiments. The larger the value, the better the performance.

\item \textbf{Mean Average Precision (MAP)} is another widely used global evaluation in ranking tasks, which measure the quality of the whole ranking list. Top-bias property of MAP is particularly significant in evaluating ranking tasks such as top-n recommendation. The larger the value, the better the performance.
\end{itemize}

\subsection{RLBL VS. TA-RLBL}

To compare the performances of our proposed RLBL and TA-RLBL, and investigate their performances with different window size, we conduct experiments on the three datasets with varying window size $n$. The results evaluated by recall, F1-score and MAP are illustrated in Table \ref{tab:mine}. We can clearly observe that, TA-RLBL performs better that RLBL in most cases. On the Movielens dataset, TA-RLBL clearly achieves a better performance evaluated by all the metrics with all the window width and the performance difference between the two models are stable. On the Global Terrorism Database, TA-RLBL performs better than RLBL mostly, especially evaluated by the global metrics MAP. But under window width $n=9$, the RLBL model achieves a slightly better recall@10 and F1-score@10 scores. These observations clearly indicate that replacing position-specific transition with time-specific transition can achieve better performance when there exists explicit time information. However, on the Tmall dataset, RLBL performs better than TA-RLBL in most cases. The reason may be that time information in the Tmall dataset is detailed to the day level. In the Tmall dataset, there are averagely $5.56$ times of clicking, $1.65$ times of purchasing, $1.42$ times of adding to favorites, and $1.25$ times of adding to shopping chart in one day conducted by one user. For these behaviors happening in the same day, there exists only sequential information of behaviors on items, but no more detailed time information. Accordingly, time-specific transition matrices for behaviors in one day will become the same, and orders among them will be discarded. Therefore, time-specific transition in TA-RLBL brings slight performance decrease on the Tmall dataset. Accordingly, it is necessary to select a proper model between RLBL and TA-RLBL according to whether there exists enough detailed time information in the dataset. For TA-RLBL incorporates time difference information, when the dataset has detailed time information, TA-RLBL will perform better. Otherwise, RLBL will be a better choice.

The experimental results in Table \ref{tab:mine} provide some hints in selecting the best window width n for RLBL and TA-RLBL in our experiments. Performances of our models on Movielens are stable and the best performances are obviously achieved at $n = 6$. On the Tmall dataset and the Global Terrorism Database, the performances are not so stable evaluated by different metrics. We can select the best parameters according to the global metric MAP, which considers all the positions in a ranking list. Then the best window width for the Global Terrorism Database is $n = 9$, and the best window width for the Tmall dataset is $n = 5$. For the rest of our experiments, we report the performances of RLBL and TA-RLBL under the best window width. Moreover, for metrics recall@p and F1-score@p, there seems existing a rough pattern. For smaller p, better recall values and F1-score values of RLBL and TA-RLBL are achieved with smaller window width n. While for larger p, better recall values and F1-score values of RLBL and TA-RLBL are achieved with larger window width $n$.

\subsection{Multiple Behaviors VS. Single Behavior}

We have analyzed performances of RLBL and TA-RLBL modeling multiple behaviors. To investigate the impact of multiple behaviors and single behavior on prediction effectiveness, we need to obtain performances of RLBL and TA-RLBL modeling a single behavior. As we ignore different types of behaviors when implementing compared methods, we also ignore multiple types of behaviors conducted on items in sequences when implementing RLBL and TA-RLBL in this experiment. Thus, the data of a user becomes a sequence consisting of items without behavioral types. Then, performances of the proposed methods under a single type of behavior can be obtained. To be noted, the partition of three datasets among training, testing and validation stays the same.

The performance comparison of modeling multiple behaviors and single behavior evaluated by recall, F1-score and MAP on three datasets is shown in Table \ref{tab:single}. We can clearly observe the significant improvements brought by modeling multiple behaviors. Comparing with modeling single behavior, MAP improvements of RLBL modeling multiple behaviors are $2.92\%$, $10.52\%$ and $6.41\%$ on three datasets respectively. And for TA-RLBL modeling multiple behaviors, comparing with modeling single behavior, the MAP improvements become $2.96\%$, $10.46\%$ and $6.42\%$, which are close to previous ones. Moreover, we can also see that, even ignoring multiple types of behaviors, RLBL and TA-RLBL can still outperform RNN with a relatively significant advantage, which indicates the effectiveness of position-specific and time-specific transition. Meanwhile, comparing results of RLBL and TA-RLBL in Table \ref{tab:mine} with results of RNN in Table \ref{tab:single}, even not with the best window width, most of the results of RLBL and TA-RLBL are still better than the performance of RNN. This indicates the effectiveness and stability of RLBL and TA-RLBL with varying window width.

\subsection{Performance Comparison with Different Methods}

\begin{table}[tb]
\caption{Comparison of multiple behaviors and single behavior.}
\centering
\subtable[Performance on the Movielens dataset with dimensionality $d=8$ and window width $n=6$.]{
    \begin{tabular}{cccccc}
    \toprule
    behaviors & method & recall@1 & recall@5 & recall@10 & MAP \\
    \midrule
    \multirow{3}[0]{*}{single} & RNN   & 0.0063  & 0.0318  & 0.0484  & 0.0362  \\
          & RLBL  & 0.0068  & 0.0343  & 0.0519  & 0.0384  \\
          & TA-RLBL & 0.0068  & 0.0360  & 0.0535  & 0.0392  \\
    \midrule
    \multirow{2}[0]{*}{multiple} & RLBL  & 0.0071  & 0.0354  & 0.0538  & 0.0395  \\
          & TA-RLBL & \textbf{0.0072} & \textbf{0.0372} & \textbf{0.0554} & \textbf{0.0404} \\
    \bottomrule
    \end{tabular}%
  \label{tab:single_movielens}%
}
\qquad
\subtable[Performance on the Global Terrorism Database with dimensionality $d=8$ and window width $n=9$.]{
    \begin{tabular}{cccccc}
    \toprule
    behaviors & method & recall@1 & recall@5 & recall@10 & MAP \\
    \midrule
    \multirow{3}[0]{*}{single} & RNN   & 0.1216  & 0.4168  & 0.5912  & 0.2600  \\
          & RLBL  & 0.1254  & 0.4723  & 0.6665  & 0.2853  \\
          & TA-RLBL & 0.1298  & 0.4783  & 0.6648  & 0.2896  \\
    \midrule
    \multirow{2}[0]{*}{multiple} & RLBL  & 0.1580  & 0.4865  & \textbf{0.6748} & 0.3153  \\
          & TA-RLBL & \textbf{0.1634} & \textbf{0.4926} & 0.6730  & \textbf{0.3199} \\
    \bottomrule
    \end{tabular}%
  \label{tab:single_GTD}%
}
\qquad
\subtable[Performance on the Tmall dataset with dimensionality $d=8$ and window width $n=5$.]{
    \begin{tabular}{cccccc}
    \toprule
    behaviors & method & recall@1 & recall@5 & recall@10 & MAP \\
    \midrule
    \multirow{3}[0]{*}{single} & RNN   & 0.1283  & 0.3410  & 0.4397  & 0.2432  \\
          & RLBL  & 0.1389  & 0.3581  & 0.5277  & 0.2666  \\
          & TA-RLBL & 0.1227  & 0.3824  & 0.5221  & 0.2636  \\
    \midrule
    \multirow{2}[0]{*}{multiple} & RLBL  & \textbf{0.1600} & 0.3822  & \textbf{0.5519} & \textbf{0.2836} \\
          & TA-RLBL & 0.1413  & \textbf{0.4081} & 0.5461  & 0.2804  \\
    \bottomrule
    \end{tabular}%
  \label{tab:single_tmall}%
}
\label{tab:single}%
\end{table}

\begin{figure}[tb]
\centering
\includegraphics[width=0.5\textwidth]{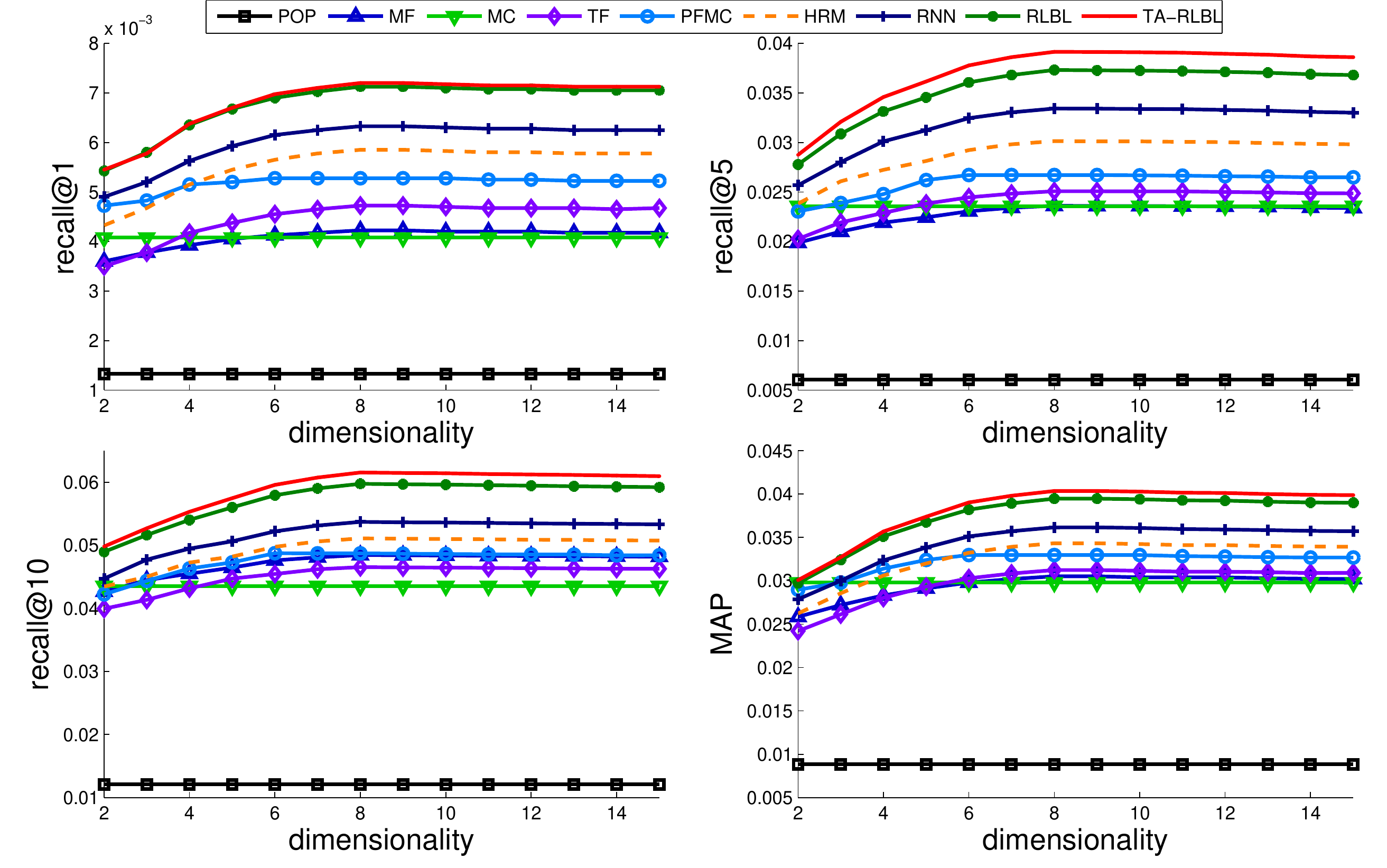}
\caption{Performance comparison on the Movielens dataset with varying dimensionality $d$ and window width $n=6$.}
\label{fig:dimen3}
\end{figure}

\begin{figure}[tb]
\centering
\includegraphics[width=0.5\textwidth]{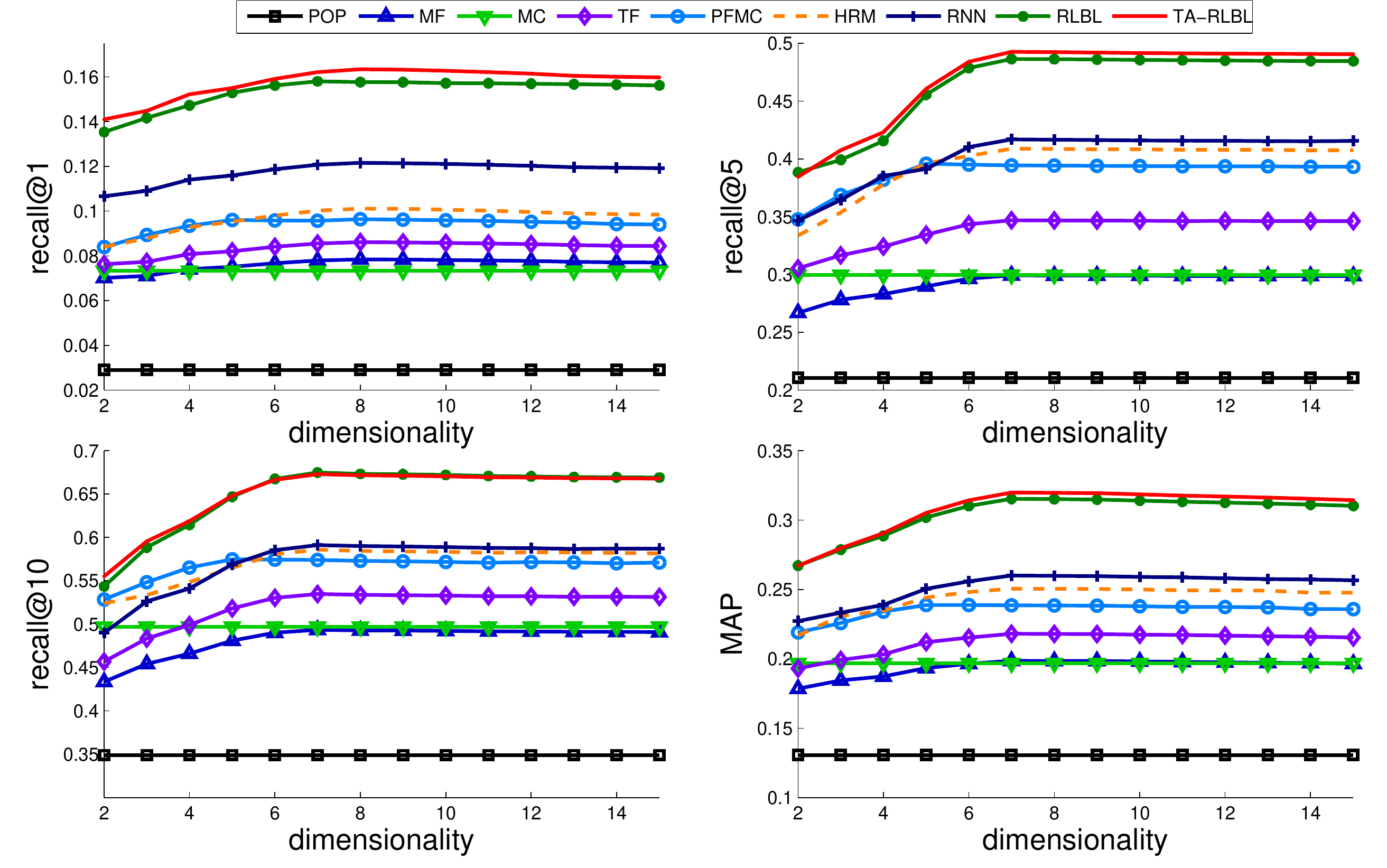}
\caption{Performance comparison on the Global Terrorism Database with varying dimensionality $d$ and window width $n=9$.}
\label{fig:dimen2}
\end{figure}

\begin{figure}[tb]
\centering
\includegraphics[width=0.5\textwidth]{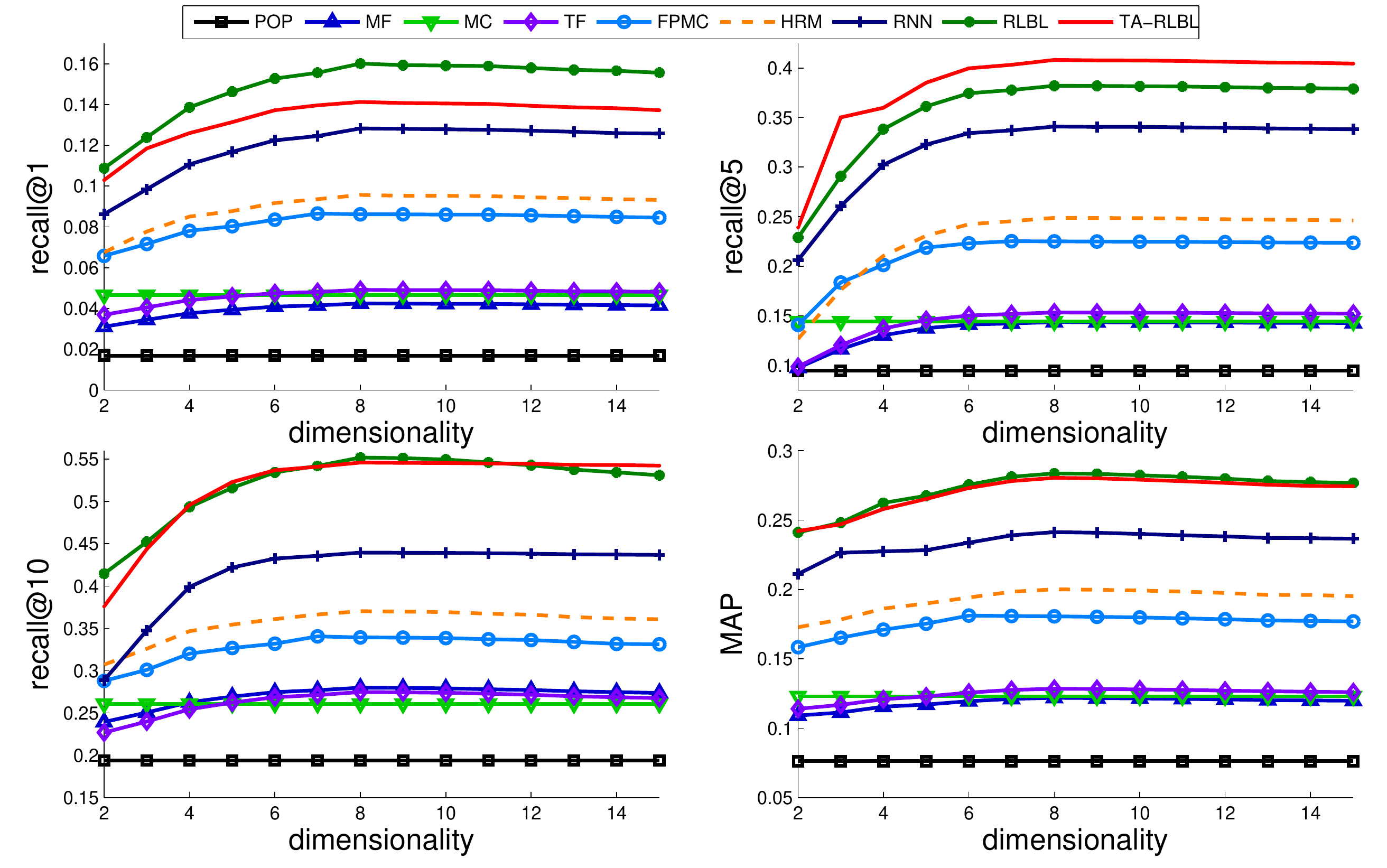}
\caption{Performance comparison on the Tmall dataset with varying dimensionality $d$ and window width $n=5$.}
\label{fig:dimen1}
\end{figure}

We compare RLBL, TA-RLBL and competitive methods with varying dimensionality $d$ evaluated by recall and MAP on the three datasets. The results on the Movielens dataset, the Global Terrorism Database and the Tmall dataset are illustrated in Figure \ref{fig:dimen3}, \ref{fig:dimen2} and \ref{fig:dimen1} respectively. Compared to the baseline performance of POP, the performances of MF, MC and TF have very similar improvement on the three datasets. They all have their shortcomings. Since MF cannot model sequential information, MC cannot model collaborative information, and TF has difficulty in predicting future behaviors, none of them achieves very satisfactory results. Jointly modeling sequential information and collaborative information, FPMC achieves great improvement comparing with these three methods. Learning latent representations of recent behaviors, HRM further improves the performance of FPMC. Furthermore, RNN brings another large improvement on the three datasets, and is clearly the best one among all the compared methods. Moreover, we can observe that, our proposed RLBL model and TA-RLBL model achieve the best performance on all the three datasets in terms of all the metrics. Using the performances with the best dimensionality of each method, comparing with RNN, the MAP improvements of RLBL are $9.18\%$, $21.27\%$ and $\%16.64$ on the Movielens dataset, the Global Terrorism Database and the Tmall dataset respectively. And the MAP improvements of TA-RLBL are $11.62\%$, $23.04\%$ and $15.31\%$ on the three datasets respectively. These results show the superiority of our methods brought by multi-behavior modeling and incorporating position-specific in RLBL or time-specific transition in TA-RLBL.

In Figure \ref{fig:dimen3}, \ref{fig:dimen2} and \ref{fig:dimen1}, we can also observe the performance curves of all the methods growing along with dimensionality $n$. All the curves clearly show the great advantages of RLBL and TA-RLBL comparing with other compared methods with different dimensionality. The performance difference between RLBL and TA-RLBL discussed above can also be observed from the performance curves. Moreover, the curves show that the performances of our models are stable in a large range on different datasets evaluated by different metrics. And even not with the best dimensionality, our methods can still outperform compared methods. According to the curves, we select the dimensionality as $d=8$, and we report corresponding performances in the rest of our experiments.

\subsection{Comparison with Different Length of Behavioral History}

Similar to the strategy in \cite{wang2015learning}, we split behavior sequences into three different types according to their length: short, medium and long. Thus, we can investigate the performance of models under different situations. In our experiments, for roughly equal splitting of behavioral sequences, we set the thresholds for the Movielens dataset as $50$ and $200$, the thresholds for the Global Terrorism Database as $50$ and $200$, and the thresholds for the Tmall dataset as $100$ and $500$.

The performance comparison of FPMC, HRM, RNN, RLBL and TA-RLBL with different length of behavioral history evaluated by recall, F1-score and MAP is shown in Table \ref{tab:length}. From the results, we can see that RLBL and TA-RLBL performs better than compared methods, i.e., FPMC, HRM and RNN, in all the situations. This shows the flexibility of our methods with variety length of behavioral history. Moreover, FPMC and HRM have the best performances on medium-length sequences, followed by long-length sequences. This also confirms the results and consequences in \cite{wang2015learning}, where FPMC and HRM perform best on medium-length sequences. For RNN, RLBL and TA-RLBL, the longer the sequences, the better the performances. This may because FPMC and HRM only model the most recent behaviors when making prediction, and previous behaviors can only be revealed by constant user latent vectors. Then, except most recent behaviors, other behaviors after model training will be ignored. So, with longer behavioral sequences, there will be more behaviors ignored, and poorer performances will be achieved. While models with recurrent structure, i.e., RNN, RLBL and TA-RLBL, can take the whole sequence into consideration, and user representations can change dynamically along with behavioral sequences. Thus, our RLBL and TA-RLBL can easily deal with the situation when sequences are too long.

\begin{table}[!tb]
\caption{Performance comparison with different behavioral history length.}
\centering
\subtable[Performance on the Movielens dataset with dimensionality $d=8$ and window width $n=6$.]{
    \begin{tabular}{cccrrrc}
    \toprule
    length & \multicolumn{2}{c}{method} & recall@1 & recall@5 & recall@10 & MAP \\
    \midrule
    \multirow{5}[0]{*}{short} & \multicolumn{2}{c}{FPMC} & \multicolumn{1}{c}{0.0052 } & \multicolumn{1}{c}{0.0250 } & \multicolumn{1}{c}{0.0433 } & 0.0325  \\
          & \multicolumn{2}{c}{HRM} & \multicolumn{1}{c}{0.0057 } & \multicolumn{1}{c}{0.0283 } & \multicolumn{1}{c}{0.0456 } & 0.0339  \\
          & \multicolumn{2}{c}{RNN} & \multicolumn{1}{c}{0.0062 } & \multicolumn{1}{c}{0.0313 } & \multicolumn{1}{c}{0.0478 } & 0.0357  \\
          & \multicolumn{2}{c}{RLBL} & \multicolumn{1}{c}{0.0070 } & \multicolumn{1}{c}{0.0351 } & \multicolumn{1}{c}{0.0535 } & 0.0391  \\
          & \multicolumn{2}{c}{TA-RLBL} & \multicolumn{1}{c}{\textbf{0.0071}} & \multicolumn{1}{c}{\textbf{0.0368}} & \multicolumn{1}{c}{\textbf{0.0550}} & \textbf{0.0400} \\
    \midrule
    \multirow{5}[0]{*}{medium} & \multicolumn{2}{c}{FPMC} & \multicolumn{1}{c}{0.0054 } & \multicolumn{1}{c}{0.0257 } & \multicolumn{1}{c}{0.0441 } & 0.0333  \\
          & \multicolumn{2}{c}{HRM} & \multicolumn{1}{c}{0.0060 } & \multicolumn{1}{c}{0.0290 } & \multicolumn{1}{c}{0.0464 } & 0.0346  \\
          & \multicolumn{2}{c}{RNN} & \multicolumn{1}{c}{0.0063 } & \multicolumn{1}{c}{0.0317 } & \multicolumn{1}{c}{0.0483 } & 0.0361  \\
          & \multicolumn{2}{c}{RLBL} & \multicolumn{1}{c}{0.0072 } & \multicolumn{1}{c}{0.0356 } & \multicolumn{1}{c}{0.0540 } & 0.0397  \\
          & \multicolumn{2}{c}{TA-RLBL} & \multicolumn{1}{c}{\textbf{0.0073}} & \multicolumn{1}{c}{\textbf{0.0373}} & \multicolumn{1}{c}{\textbf{0.0556}} & \textbf{0.0405} \\
    \midrule
    \multirow{5}[0]{*}{long} & \multicolumn{2}{c}{FPMC} & \multicolumn{1}{c}{0.0053 } & \multicolumn{1}{c}{0.0254 } & \multicolumn{1}{c}{0.0438 } & 0.0330  \\
          & \multicolumn{2}{c}{HRM} & \multicolumn{1}{c}{0.0059 } & \multicolumn{1}{c}{0.0287 } & \multicolumn{1}{c}{0.0460 } & 0.0344  \\
          & \multicolumn{2}{c}{RNN} & \multicolumn{1}{c}{0.0064 } & \multicolumn{1}{c}{0.0320 } & \multicolumn{1}{c}{0.0487 } & 0.0364  \\
          & \multicolumn{2}{c}{RLBL} & \multicolumn{1}{c}{0.0073 } & \multicolumn{1}{c}{0.0359 } & \multicolumn{1}{c}{0.0544 } & 0.0400  \\
          & \multicolumn{2}{c}{TA-RLBL} & \multicolumn{1}{c}{\textbf{0.0074}} & \multicolumn{1}{c}{\textbf{0.0377}} & \multicolumn{1}{c}{\textbf{0.0561}} & \textbf{0.0409} \\
    \bottomrule
    \end{tabular}%
  \label{tab:length_movie}%
}
\qquad
\subtable[Performance on the Global Terrorism Database with dimensionality $d=8$ and window width $n=9$.]{
    \begin{tabular}{cccrrrc}
    \toprule
    length & \multicolumn{2}{c}{method} & recall@1 & recall@5 & recall@10 & MAP \\
    \midrule
    \multirow{5}[0]{*}{short} & \multicolumn{2}{c}{FPMC} & \multicolumn{1}{c}{0.0935 } & \multicolumn{1}{c}{0.3834 } & \multicolumn{1}{c}{0.5658 } & 0.2341  \\
          & \multicolumn{2}{c}{HRM} & \multicolumn{1}{c}{0.0966 } & \multicolumn{1}{c}{0.3980 } & \multicolumn{1}{c}{0.5725 } & 0.2410  \\
          & \multicolumn{2}{c}{RNN} & \multicolumn{1}{c}{0.1180 } & \multicolumn{1}{c}{0.4066 } & \multicolumn{1}{c}{0.5833 } & 0.2544  \\
          & \multicolumn{2}{c}{RLBL} & \multicolumn{1}{c}{0.1507 } & \multicolumn{1}{c}{0.4728 } & \multicolumn{1}{c}{\textbf{0.6639}} & 0.3073  \\
          & \multicolumn{2}{c}{TA-RLBL} & \multicolumn{1}{c}{\textbf{0.1557}} & \multicolumn{1}{c}{\textbf{0.4770}} & \multicolumn{1}{c}{0.6621 } & \textbf{0.3124} \\
    \midrule
    \multirow{5}[0]{*}{medium} & \multicolumn{2}{c}{FPMC} & \multicolumn{1}{c}{0.0981 } & \multicolumn{1}{c}{0.4006 } & \multicolumn{1}{c}{0.5748 } & 0.2422  \\
          & \multicolumn{2}{c}{HRM} & \multicolumn{1}{c}{0.1029 } & \multicolumn{1}{c}{0.4124 } & \multicolumn{1}{c}{0.5830 } & 0.2503  \\
          & \multicolumn{2}{c}{RNN} & \multicolumn{1}{c}{0.1216 } & \multicolumn{1}{c}{0.4168 } & \multicolumn{1}{c}{0.5912 } & 0.2600  \\
          & \multicolumn{2}{c}{RLBL} & \multicolumn{1}{c}{0.1567 } & \multicolumn{1}{c}{0.4840 } & \multicolumn{1}{c}{\textbf{0.6734}} & 0.3140  \\
          & \multicolumn{2}{c}{TA-RLBL} & \multicolumn{1}{c}{\textbf{0.1620}} & \multicolumn{1}{c}{\textbf{0.4906}} & \multicolumn{1}{c}{0.6710 } & \textbf{0.3183} \\
    \midrule
    \multirow{5}[0]{*}{long} & \multicolumn{2}{c}{FPMC} & \multicolumn{1}{c}{0.0964 } & \multicolumn{1}{c}{0.3944 } & \multicolumn{1}{c}{0.5741 } & 0.2385  \\
          & \multicolumn{2}{c}{HRM} & \multicolumn{1}{c}{0.1007 } & \multicolumn{1}{c}{0.4068 } & \multicolumn{1}{c}{0.5824 } & 0.2468  \\
          & \multicolumn{2}{c}{RNN} & \multicolumn{1}{c}{0.1239 } & \multicolumn{1}{c}{0.4233 } & \multicolumn{1}{c}{0.5918 } & 0.2642  \\
          & \multicolumn{2}{c}{RLBL} & \multicolumn{1}{c}{0.1599 } & \multicolumn{1}{c}{0.4916 } & \multicolumn{1}{c}{\textbf{0.6790}} & 0.3178  \\
          & \multicolumn{2}{c}{TA-RLBL} & \multicolumn{1}{c}{\textbf{0.1649}} & \multicolumn{1}{c}{\textbf{0.4985}} & \multicolumn{1}{c}{0.6766 } & \textbf{0.3227} \\
    \bottomrule
    \end{tabular}%
  \label{tab:length_GTD}%
}
\qquad
\subtable[Performance on the Tmall dataset with dimensionality $d=8$ and window width $n=5$.]{
    \begin{tabular}{cccrrrc}
    \toprule
    length & \multicolumn{2}{c}{method} & recall@1 & recall@5 & recall@10 & MAP \\
    \midrule
    \multirow{5}[0]{*}{short} & \multicolumn{2}{c}{FPMC} & \multicolumn{1}{c}{0.0837 } & \multicolumn{1}{c}{0.2330 } & \multicolumn{1}{c}{0.3350 } & 0.1807  \\
          & \multicolumn{2}{c}{HRM} & \multicolumn{1}{c}{0.0934 } & \multicolumn{1}{c}{0.2588 } & \multicolumn{1}{c}{0.3668 } & 0.1990  \\
          & \multicolumn{2}{c}{RNN} & \multicolumn{1}{c}{0.1251 } & \multicolumn{1}{c}{0.3363 } & \multicolumn{1}{c}{0.4350 } & 0.2401  \\
          & \multicolumn{2}{c}{RLBL} & \multicolumn{1}{c}{\textbf{0.1566}} & \multicolumn{1}{c}{0.3786 } & \multicolumn{1}{c}{\textbf{0.5494}} & \textbf{0.2811} \\
          & \multicolumn{2}{c}{TA-RLBL} & \multicolumn{1}{c}{0.1381 } & \multicolumn{1}{c}{\textbf{0.4041}} & \multicolumn{1}{c}{0.5432 } & 0.2780  \\
    \midrule
    \multirow{5}[0]{*}{medium} & \multicolumn{2}{c}{FPMC} & \multicolumn{1}{c}{0.0872 } & \multicolumn{1}{c}{0.2393 } & \multicolumn{1}{c}{0.3412 } & 0.1848  \\
          & \multicolumn{2}{c}{HRM} & \multicolumn{1}{c}{0.0971 } & \multicolumn{1}{c}{0.2653 } & \multicolumn{1}{c}{0.3734 } & 0.2032  \\
          & \multicolumn{2}{c}{RNN} & \multicolumn{1}{c}{0.1282 } & \multicolumn{1}{c}{0.3410 } & \multicolumn{1}{c}{0.4396 } & 0.2432  \\
          & \multicolumn{2}{c}{RLBL} & \multicolumn{1}{c}{\textbf{0.1608}} & \multicolumn{1}{c}{0.3841 } & \multicolumn{1}{c}{\textbf{0.5547}} & \textbf{0.2850} \\
          & \multicolumn{2}{c}{TA-RLBL} & \multicolumn{1}{c}{0.1420 } & \multicolumn{1}{c}{\textbf{0.4102}} & \multicolumn{1}{c}{0.5491 } & 0.2818  \\
    \midrule
    \multirow{5}[0]{*}{long} & \multicolumn{2}{c}{FPMC} & \multicolumn{1}{c}{0.0859 } & \multicolumn{1}{c}{0.2369 } & \multicolumn{1}{c}{0.3387 } & 0.1831  \\
          & \multicolumn{2}{c}{HRM} & \multicolumn{1}{c}{0.0957 } & \multicolumn{1}{c}{0.2627 } & \multicolumn{1}{c}{0.3703 } & 0.2016  \\
          & \multicolumn{2}{c}{RNN} & \multicolumn{1}{c}{0.1303 } & \multicolumn{1}{c}{0.3445 } & \multicolumn{1}{c}{0.4433 } & 0.2447  \\
          & \multicolumn{2}{c}{RLBL} & \multicolumn{1}{c}{\textbf{0.1633}} & \multicolumn{1}{c}{0.3879 } & \multicolumn{1}{c}{\textbf{0.5588}} & \textbf{0.2876} \\
          & \multicolumn{2}{c}{TA-RLBL} & \multicolumn{1}{c}{0.1439 } & \multicolumn{1}{c}{\textbf{0.4143}} & \multicolumn{1}{c}{0.5532 } & 0.2842  \\
    \bottomrule
    \end{tabular}%
  \label{tab:length_tmall}%
}
\label{tab:length}%
\end{table}

\subsection{Analysis on Computational Time and Convergence}

To investigate the efficiency of our proposed methods, we illustrate the computational time of RNN, RLBL and TA-RLBL in each iteration during training on three datasets in Table \ref{tab:runtime}. The computation time is measured in seconds. Here, according to previous experimental results, the dimensionality is chosen to be $d=8$. And the window width is $n=6$, $n=9$ and $n=5$ on Movielens, GTD and Tmall respectively. Experiments are conducted on a computer with an 8 core 3.0 GHz CPU, 16 GB RAM, and a NVIDIA TITAN X GPU. From results in Table \ref{tab:runtime}, we can observe that all three methods can be trained in an acceptable time. RLBL is a little faster than TA-RLBL, indicating that time-specific transition is a little more time consuming than position-specific transition. Moreover, the computational time of RLBL and TA-RLBL is less than twice of that of conventional RNN. This means that, the significant performance improvement brought by RLBL and TA-RLBL only requires no more than double computational time.

\begin{table}[!tb]
  \centering
  \caption{The computational time of RNN, RLBL and TA-RLBL in each iteration during training on three datasets.}
    \begin{tabular}{cccc}
    \toprule
    method & Movielens & GTD   & Tmall \\
    \midrule
    RNN   & 902s  & 115s  & 335s \\
    RLBL  & 1628s & 196s  & 638s \\
    TA-RLBL & 1664s & 210s  & 668s \\
    \bottomrule
    \end{tabular}%
  \label{tab:runtime}%
\end{table}%

\begin{figure*}[!tb]
\centering
\subfigure[Movielens.]{
\begin{minipage}[b]{0.31\textwidth}
\includegraphics[width=1\textwidth]{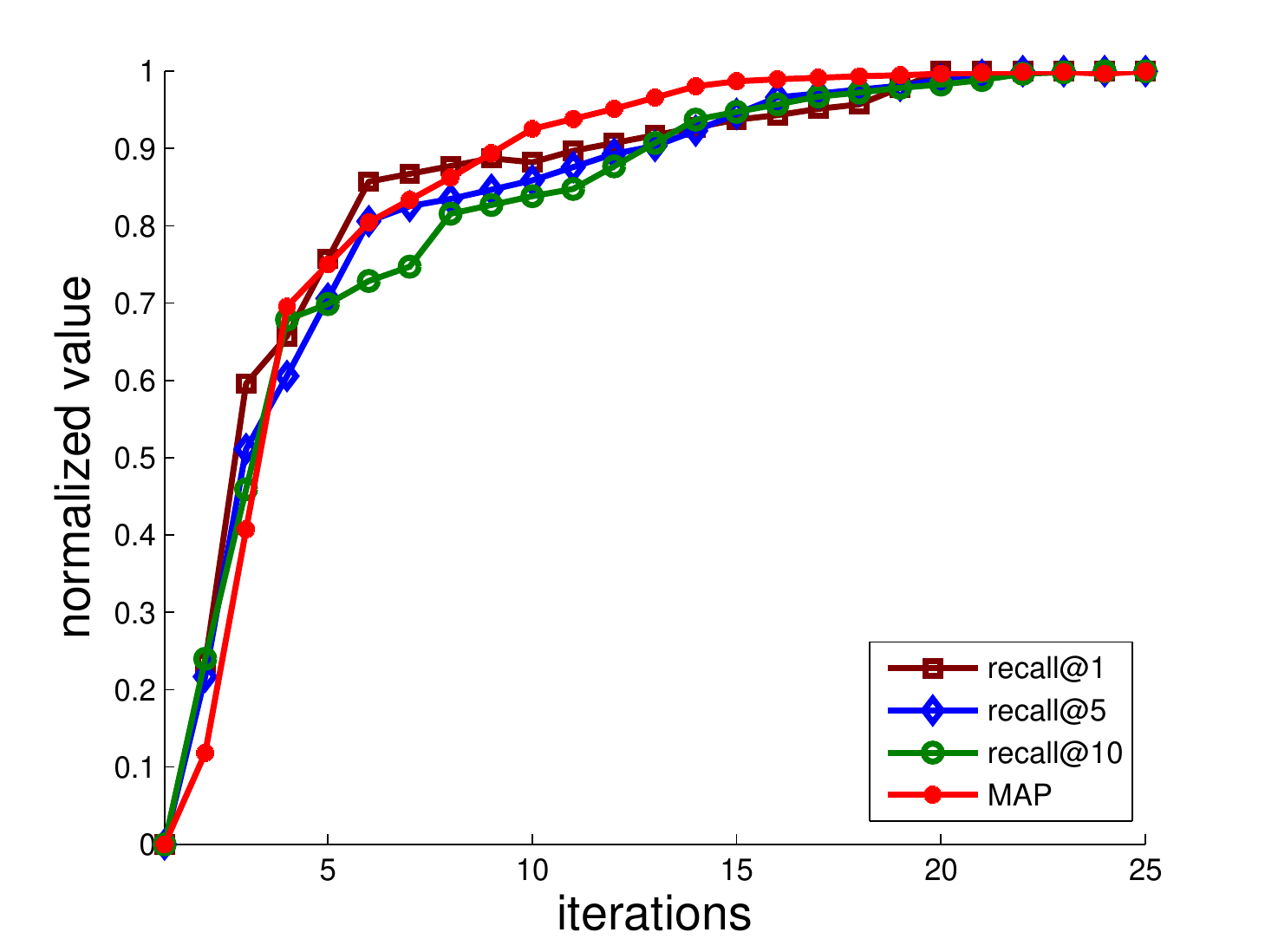}
\label{conve1_movie}
\end{minipage}
}
\subfigure[GTD.]{
\begin{minipage}[b]{0.31\textwidth}
\includegraphics[width=1\textwidth]{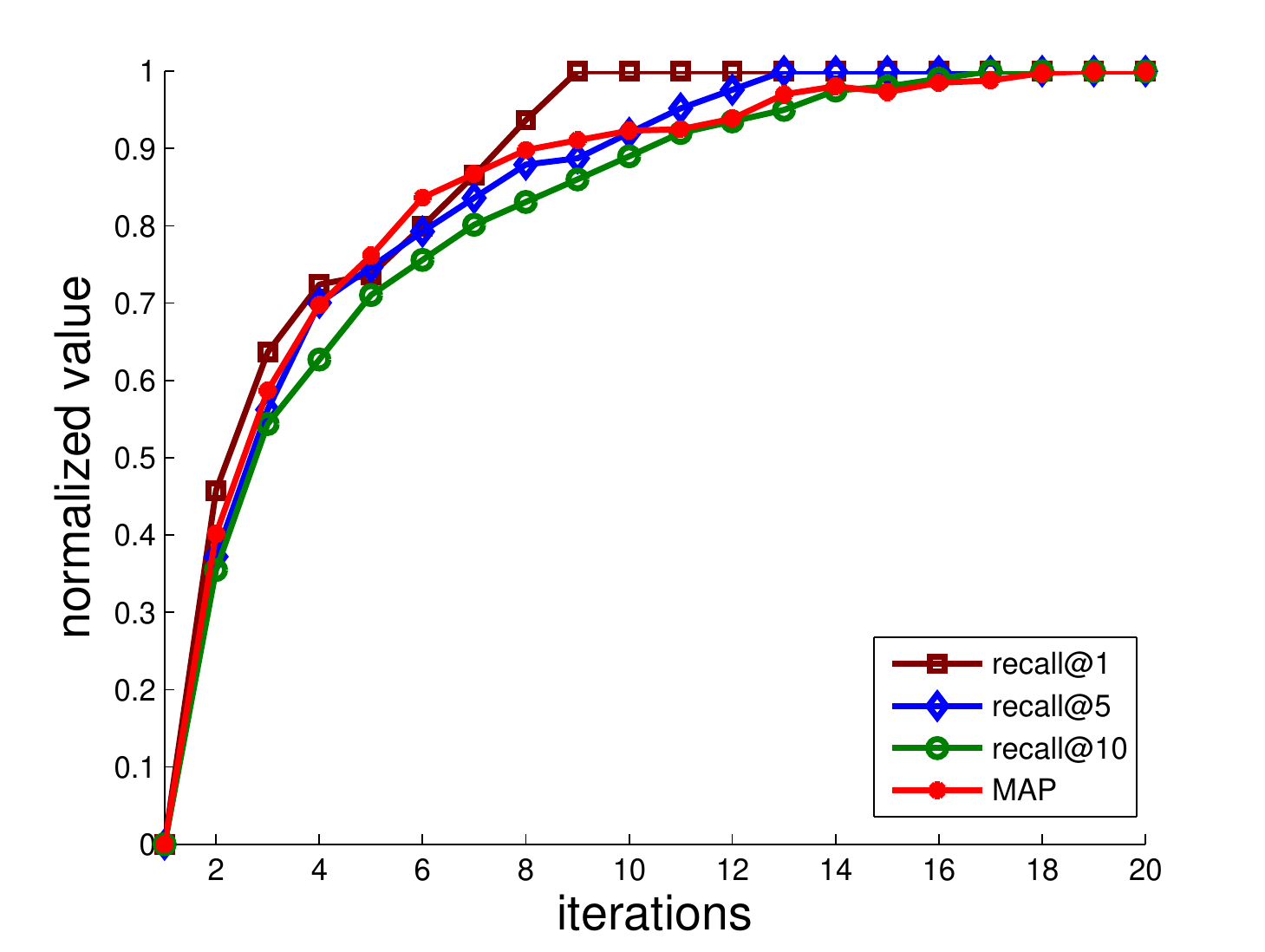}
\label{conve1_GTD}
\end{minipage}
}
\subfigure[Tmall.]{
\begin{minipage}[b]{0.31\textwidth}
\includegraphics[width=1\textwidth]{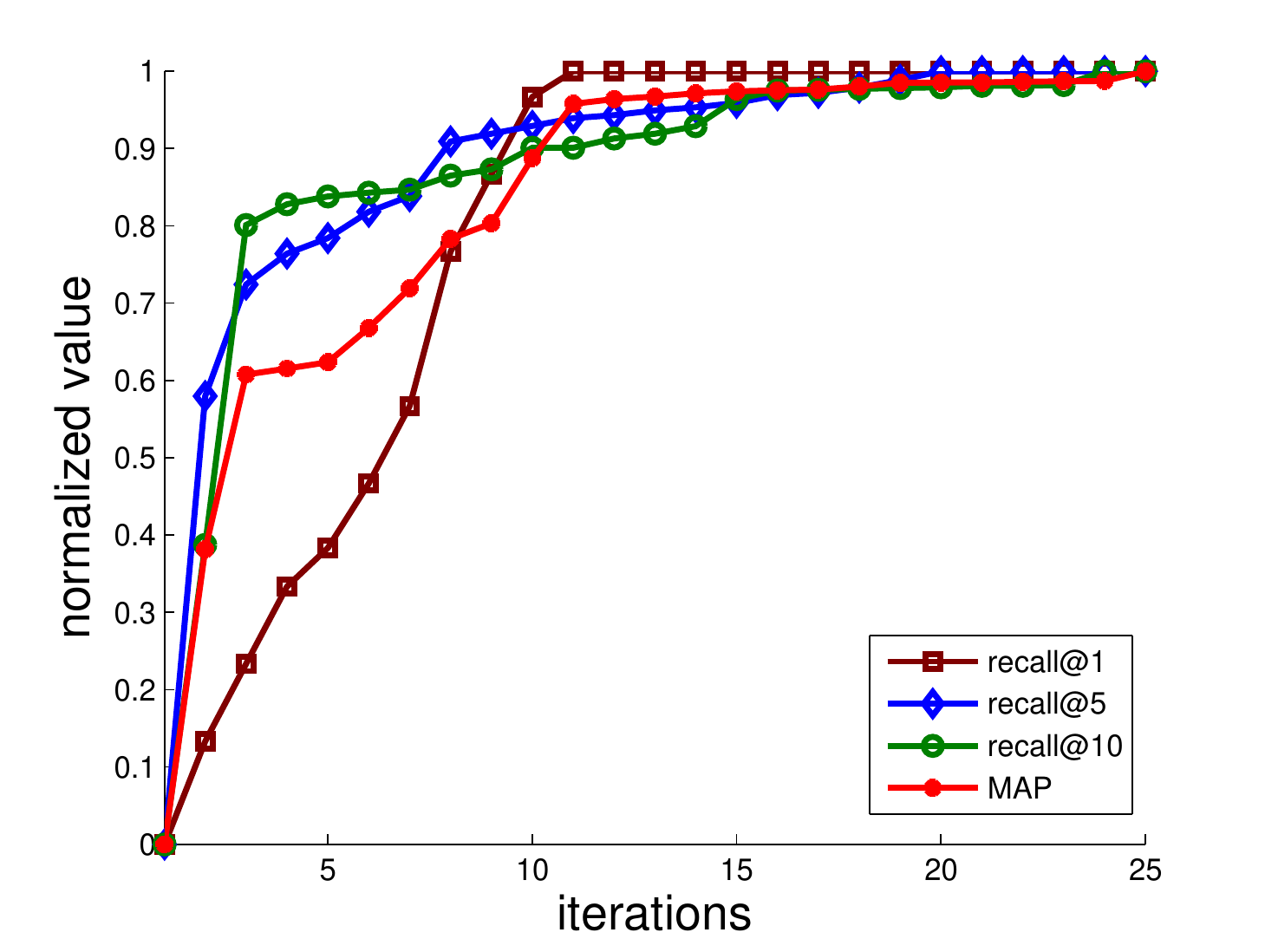}
\label{conve1_taobao}
\end{minipage}
}
\caption{Convergence curves of RLBL measured by normalized recall and MAP.}
\label{fig:conve1}
\end{figure*}

\begin{figure*}[!tb]
\centering
\subfigure[Movielens.]{
\begin{minipage}[b]{0.31\textwidth}
\includegraphics[width=1\textwidth]{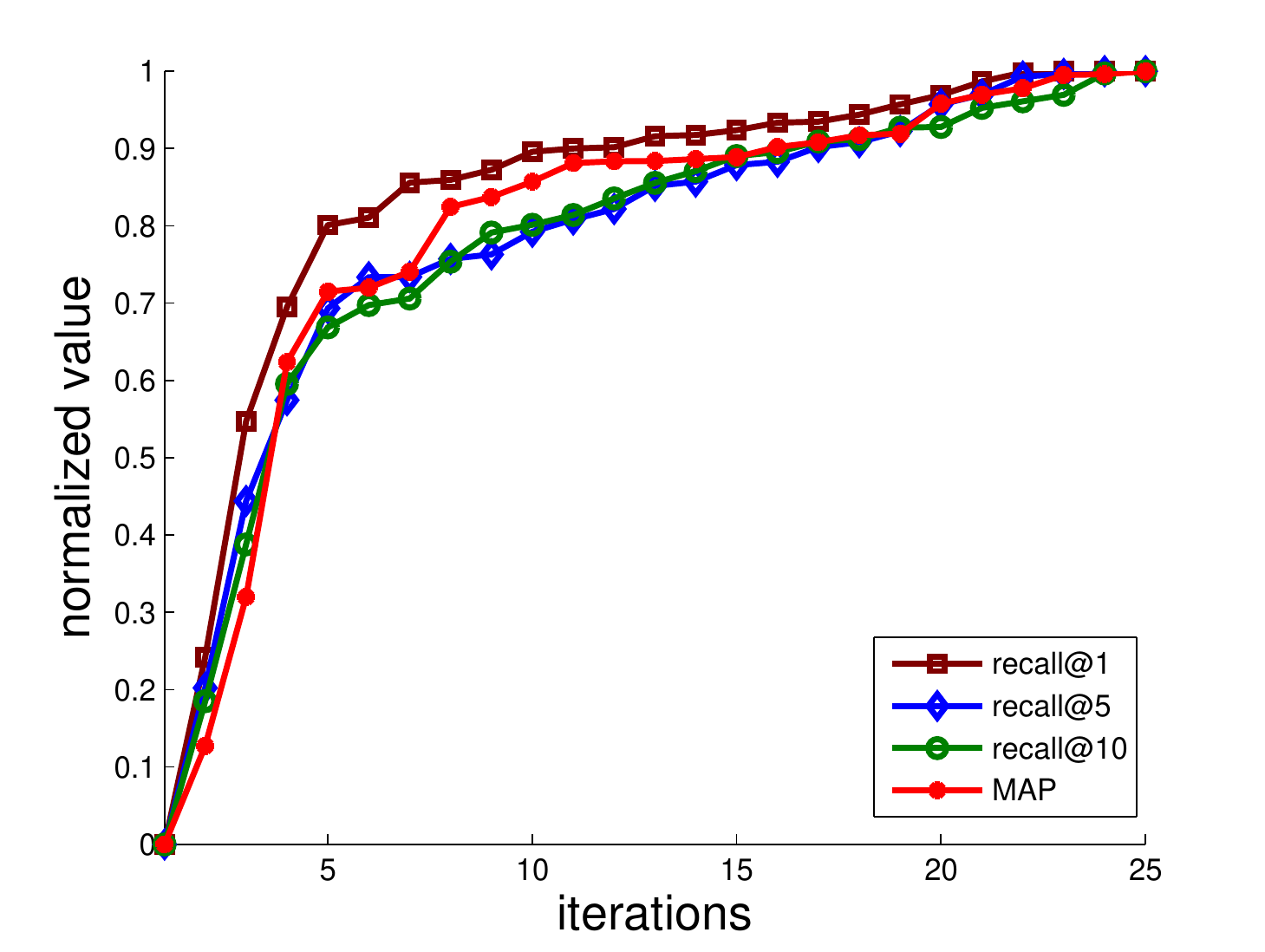}
\label{conve2_movie}
\end{minipage}
}
\subfigure[GTD.]{
\begin{minipage}[b]{0.31\textwidth}
\includegraphics[width=1\textwidth]{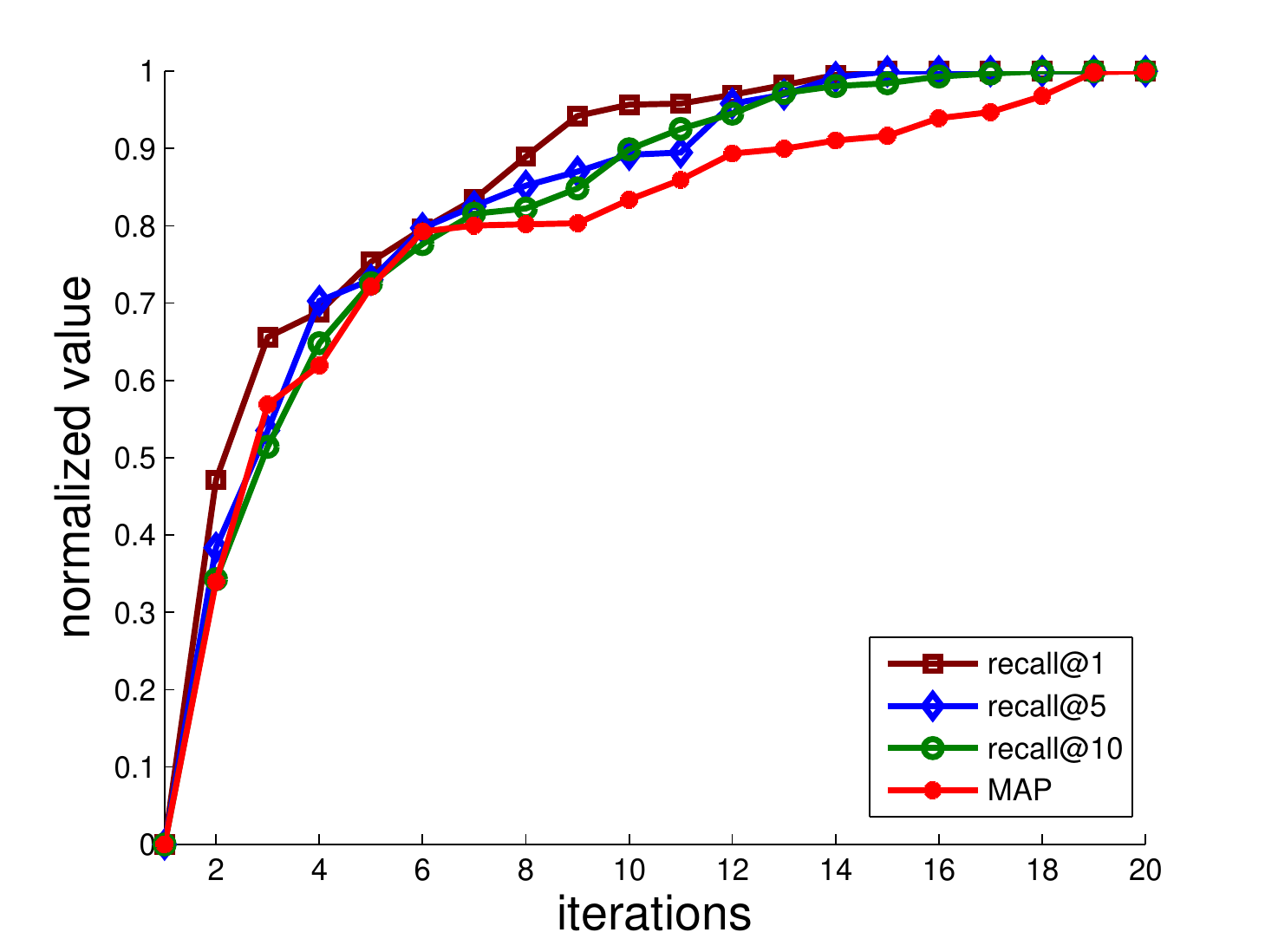}
\label{conve2_GTD}
\end{minipage}
}
\subfigure[Tmall.]{
\begin{minipage}[b]{0.31\textwidth}
\includegraphics[width=1\textwidth]{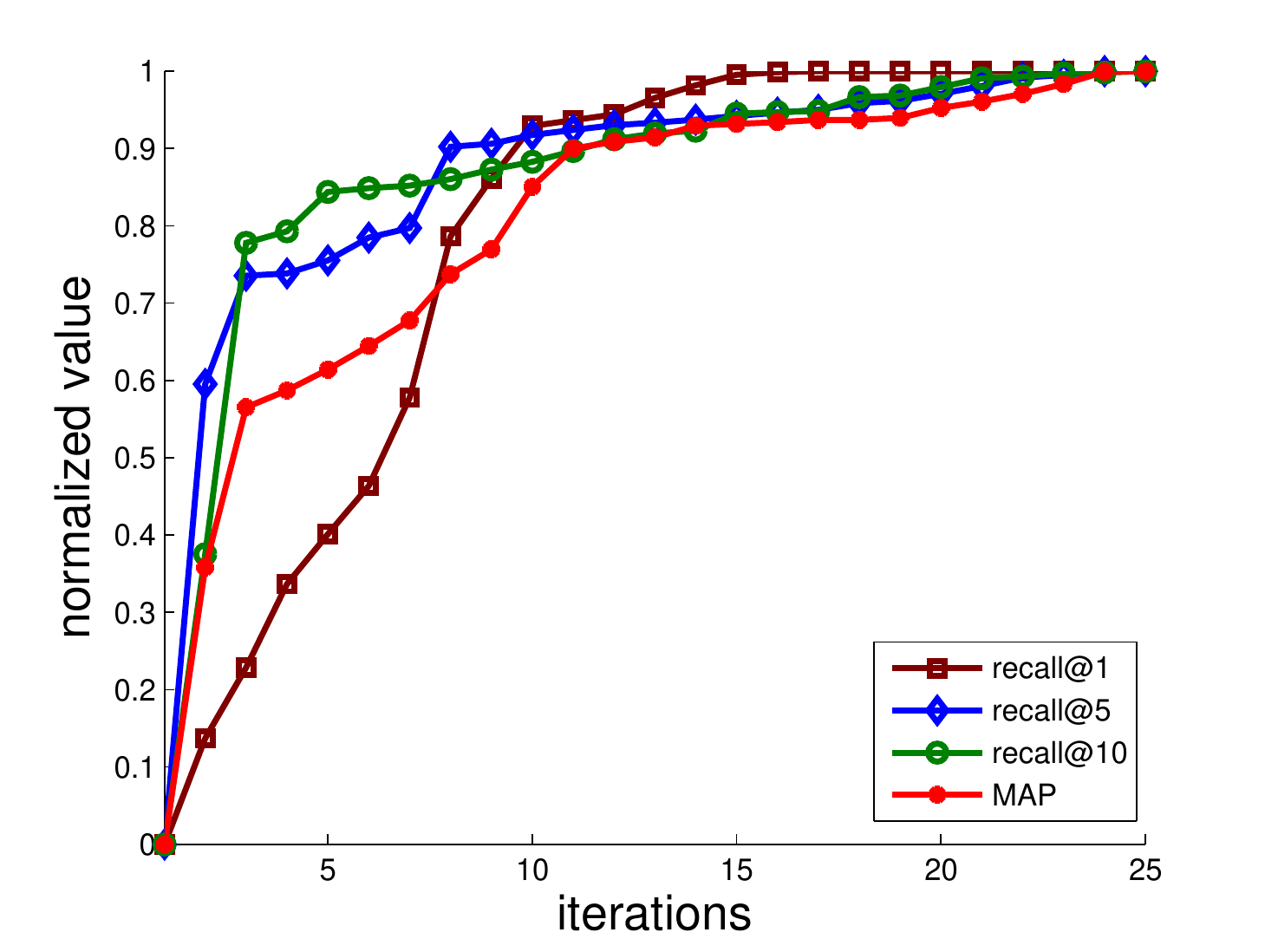}
\label{conve2_taobao}
\end{minipage}
}
\caption{Convergence curves of TA-RLBL measured by normalized recall and MAP.}
\label{fig:conve2}
\end{figure*}

Moreover, we illustrate the convergence curves of RLBL and TA-RLBL in Figure \ref{fig:conve1} and \ref{fig:conve2} respectively. To illustrate curves measured by different evaluation metrics in one figure, we calculate normalized recall and MAP of RLBL and TA-RLBL on three datasets. We normalize the values of recall and MAP into $[0,1]$, and illustrate the corresponding convergence curves. From the convergence curves, we can observe that, both RLBL and TA-RLBL can achieve convergence in a relatively small number of iterations. Moreover, recall@1 values achieves convergence faster than recall@5 values, and recall@5 values achieve convergence faster than recall@10 values. This may indicate that, the more items generated in the ranking list, the more iterations are needed during training.

\section{Conclusions and Future Work}

In this paper, we have proposed two novel multi-behavioral sequential prediction methods, i.e. recurrent log-bilinear model and time-aware recurrent log-bilinear model. We build our model under a recurrent structure. RLBL models several elements in each hidden layer and incorporate position-specific transition matrices. With such architecture, RLBL can well model both short- and long-term contexts in a historical sequence. Besides, to capture multiple types of behavior in behavioral sequences, behavior-specific matrices are designed and applied for each type of behavior. Then, to incorporate time difference information in behavioral sequences, we further extend the RLBL model and propose a time-aware recurrent log-bilinear model with time-specific transition matrices. Modeling time difference information, TA-RLBL can further improves the performance of RLBL in sequential prediction. The experimental results on three real datasets show that both RLBL and TA-RLBL outperforms state-of-the-art sequential prediction models.

In the future, we can further investigate the following direction. In RLBL and TA-RLBL, transition matrices are the same for different users, which does not confirm to practical situations. So, we need to find a method to determine different transition matrices for different users or different user groups. Moreover, we didn't take items' features, e.g., categories, descriptions and images of items, into consideration. Thus, incorporating RLBL and TA-RLBL with features of items may also be our next step.

\balance

\bibliographystyle{abbrv}
\bibliography{IEEE}

\end{document}